\documentclass{article}

\usepackage{a4wide}
\usepackage{amsmath,amssymb,amsthm}
\usepackage{color}
\usepackage[all]{xy}
\usepackage{mathtools}
\usepackage{url}

\usepackage{graphicx}
\usepackage[usenames,dvipsnames]{pstricks}
\usepackage{epsfig}
\usepackage{pst-grad}
\usepackage{pst-plot}
\usepackage{subcaption}

\usepackage{siunitx}
\newcommand{\numD}[1]{\num[round-precision = 2]{#1}}

\newcommand{\SFcaption}[1]{\tiny{#1}}

\theoremstyle{remark}

\theoremstyle{definition}

\numberwithin{equation}{section}

\newcommand{\mdef}{\mathop{\stackrel{\mbox{\tiny def}}{=}}}
\newcommand{\myeq}[2]{\begin{equation}\label{rst:#1}#2\end{equation}}

\newcommand{\myeqAN}[1]{\begin{align*}#1\end{align*}}
\newcommand{\myref}[1]{(\ref{rst:#1})}

\newcommand{\bkR}{\mathbb{R}}

\newcommand{\TS}{\mathcal{S}}
\newcommand{\TP}{\mathcal{P}}
\newcommand{\TI}{\mathcal{I}}
\newcommand{\TL}{\mathcal{L}}
\newcommand{\TT}{\mathcal{T}}
\newcommand{\TX}{\mathcal{X}}

\begin{document}

\title{The effect of immigrant communities coming from higher incidence
tuberculosis regions to a host country\thanks{This is a preprint of a 
paper whose final and definite form is with 'Ricerche di Matematica', 
ISSN 0035-5038 (Print) 1827-3491 (Online), 
available at [\url{http://link.springer.com/journal/11587}]. 
Submitted 10-Feb-2016; Revised and Accepted 31-Jan-2017.}}

\author{Eug\'{e}nio M. Rocha\\
\texttt{eugenio@ua.pt}
\and Cristiana J. Silva\\
\texttt{cjoaosilva@ua.pt}
\and Delfim F. M. Torres\thanks{Corresponding author. Email: delfim@ua.pt}\\
\texttt{delfim@ua.pt}}

\date{Center for Research and Development in Mathematics and Applications (CIDMA)\\
Department of Mathematics, University of Aveiro, 3810--193 Aveiro, Portugal}

\maketitle

% ------------------------------------------

\begin{abstract}
We introduce a new tuberculosis (TB) mathematical model, with $25$ state-space
variables where $15$ are evolution disease states (EDSs), which generalises
previous models and takes into account the (seasonal) flux of populations between a high
incidence TB country (A) and a host country (B) with low TB incidence, where (B) 
is divided into a community (G) with high percentage of people from (A) plus the 
rest of the population (C). Contrary to some beliefs, related to the fact that
agglomerations of individuals increase proportionally to the disease spread,
analysis of the model shows that the existence of semi-closed communities are
beneficial for the TB control from a global viewpoint. The model and techniques 
proposed are applied to a case-study with concrete parameters, 
which model the situation of Angola (A) and Portugal (B),
in order to show its relevance and meaningfulness.
Simulations show that variations of the transmission
coefficient on the origin country has a big influence on the number of infected
(and infectious) individuals on the community and the host country. Moreover,
there is an optimal ratio for the distribution of individuals in (C) versus (G), which
minimizes the reproduction number $R_0$. Such value does not give the minimal
total number of infected individuals in all (B), since such is attained when
the community (G) is completely isolated (theoretical scenario).
Sensitivity analysis and curve fitting on $R_0$ and on EDSs are pursuit
in order to understand the TB effects in the global statistics, by measuring
the variability of the relevant parameters. We also show that the TB transmission 
rate $\beta$ does not act linearly on $R_0$, as is common in compartment models
where system feedback or group interactions do not occur. Further,
we find the most important parameters for the increase of each EDS.
\end{abstract}



% -----------------------------------------

\section{Introduction}

Tuberculosis (TB) is an infectious disease caused by the \emph{Mycobacterium
tuberculosis (Mtb)}. Following the World Health Organization (WHO),
the \emph{(Mtb)} is the second cause of death worldwide
from a single infectious agent, after the human immunodeficiency virus
\cite{WHO:TB:report:2013}. TB is present in all regions of the world. Most of
the estimated number of cases in 2013 occurred in Asia ($56\%$) and the African
region ($29\%$); smaller proportions of cases occurred in the Eastern
Mediterranean region ($8\%$), the European region (4\%) and the region
of the Americas ($3\%$) \cite{WHO:TB:report:2014}.

In TB spread, migration plays an important role, e.g., following the International
Organization for Migration (IOM), TB is a social disease and migration, as a social
determinant of health, increases TB-related morbidity and mortality among migrants
and surrounding communities \cite{IOM:migration:TB}. Migrants of specific legal
and social status, such as workers, undocumented migrants, trafficked and detained
persons, face particular TB vulnerabilities. Among migrant workers with a legal
status, their access to TB diagnosis and care is subject to their ability to access
health care services and health insurance coverage, provided either by the state
or the employer. Illegal migrants face particular challenges such as fear of
deportation that delay or limit their access to diagnostic and treatment services.
Deportation while on treatment or poor compliance with treatment may lead to drug
resistant infection and increased chances of spreading TB in countries of origin,
transit and destination \cite{IOM:migration:TB}.

Mathematical models are an important tool in analyzing the spread and control
of infectious diseases \cite{Hethcote:1001models,Hethcote:SIAM:Rev}. There are
many mathematical dynamic models for TB, see, e.g.,
\cite{Blower:etall:1996,Castillo:Feng:1998,Cohen:Murray:2004,Gomes:etall:JTB:2007,Vynnycky:Fine:1997}
and references cited therein. There are also models dedicated to study TB
transmission dynamics in immigrants and local population. Usually, these models
divide the total population into two subgroups: immigrants and local subpopulation.
Each subgroup is divided into several epidemiological compartments: susceptible,
latent, infectious, recovered, or other, depending on the type of the model, see, e.g.,
\cite{Mod:TB:immigration:Driessche:2001,Mod:immigrat:TB:TPB:2008,Mod:Immig:Netherland:IJTLD:2002,TB:immigration:JTB:2008}.
In general, compartment models written with ordinary differential equations tend
to be nice approximations of the true scenario that have rather simple formulation,
e.g., with five state-space variables and a (non)autonomous quadratic vector field,
because of numerical and analytic limitations and the tradeoff between complexity
and the relevant information that they can present. In particular, heterogeneous
situations may be studied using such models. However, no interaction between
individuals in the different groups are considered in such models. We are interested
in understanding how the flux and distribution of individuals affects TB on a
host country. As a case-study, we have considered the situation of Angola
and Portugal, although the techniques may be applied to any similar situation.

Angola is the seventh-largest country in Southern Africa with a total population
of approximately $24.3$ million \cite{INE2014}. WHO predicts that by 2017 the
TB cases rate may rise significantly in Angola. A natural question is to try to
understand how this may affect the rest of the world. According to Celestino
Teixeira, the Coordinator of the \emph{Fight Against Tuberculosis Programme},
in 2013 Angola reported a total of $60, 807$ cases of TB in all forms, observing
an increase of $11\%$ over the previous year \cite{news:TB:ANGOLA:increase}.
Portugal is a country in Southwest Europe with a total population of approximately
$10.5$ million  \cite{INE2014}. In 2014, for the first time, the incidence of TB
in Portugal was estimated to be lower than $20$ new cases per $100,000$ inhabitants,
placing Portugal among the countries with low TB incidence. However, there are
still some regions (Lisbon and Porto) with much higher TB incidences
\cite{relatorioTB:PT:2015}. Portugal is a relevant geographically area of study
for TB because its infection behaviour is not similar to the rest of Europe,
in the sense that has higher incidence of tuberculosis. Aside from the
independence period, Angola is characterized by a reduced emigration and is
becoming gradually an attractive region, receiving migrants from different
regions, including Portugal \cite{livro:migracao:PT:2011}. Following the
Portuguese Emigration Observatory, in 2014 there were 126,356 Portuguese
emigrants living in Angola \cite{url:obsEmigraPT}. According to
the Organisation for Economic Co-operation and Development (OECD)
\cite{OCDE:migration:PT:2014}, for the first time in five years, 2012 saw the
number of long-term entry visas grow. Visas to Angolans doubled in 2012, mainly for study.
According to the Portuguese Foreigners and Borders Service, in 2012 there were
20,177 Angolans citizens living in Portugal \cite{reportImigrants:PT:2013}.
Although Angolans living in Portugal are dispersed throughout the country,
there is a very high concentration in the district of Lisbon, followed by
Set\'{u}bal and Porto \cite{ImigAngolanosPT}.

In this paper, we propose and study a new mathematical model for TB that
generalises the one proposed in \cite{TBportugalGomesRodrigues}. We consider
three different populations: people living in a high TB incidence country (A),
people living in a low TB incidence country in a semi-closed community of the
high incidence country natives (G), and the other persons living in the low
incidence country (C). Each of these three groups of population are subdivided
into the five epidemiological categories considered in the model from
\cite{TBportugalGomesRodrigues}. Our model considers the movement of persons
from the high TB incidence country to the low TB incidence country and vice-versa.
We assume that the individuals that arrive and depart from the low TB incidence
country are split into the ones that enter/leave the semi-closed community
of the high TB incidence country natives and the ones that enter/leave other
regions of the low TB incidence country. Our model is quite different from
\cite{TBportugalGomesRodrigues} and other TB models in the literature,
since it has internal transfer of individuals between the subgroups,
high TB incidence country, semi-closed community of high TB incidence
country natives and other persons living in the low TB incidence country.
We consider a case study where the low TB incidence country is represented
by Portugal and the high TB incidence country is represented by Angola.

The paper is organized as follows. In Section~\ref{sec:model}, we explain how
we construct our model. The basic reproduction number is algebraically and
numerically computed in Section~\ref{sec:R0} for the autonomous case. This
section also includes a sensitivity analysis of the basic reproduction number
with respect to TB transmission rates, transfer of individuals and ratio
of individuals that stay in the community versus spread in the host country.
Section~\ref{sec:numeric} is devoted to numerical simulations, which help us
to make a qualitative sensitivity analysis for each epidemiological category
of the subgroups Angola, semi-closed community of Angola natives and other
persons living in Portugal, when relevant TB parameters are perturbed.
We end with Section~\ref{sec:conclusions} of conclusions and future work.

% ------------------------------------------

\section{Mathematical model}
\label{sec:model}

We construct a model with three components, based on \cite{TBportugalGomesRodrigues},
where there exists seasonal flux of population between some of the components.
The model from \cite{TBportugalGomesRodrigues} divides the
total population $N$ in five epidemiological compartments: susceptible individuals
($\TS$) that never have been in contact with \emph{(Mtb)}, primary infected individuals
($\TP$) that have been infected by \emph{(Mtb)} but it is not certain if the disease
will progress, actively infected and infectious individuals ($\TI$) that are
not yet in treatment, latent infected individuals ($\TL$) and under treatment
individuals ($\TT$). Susceptible individuals become primary infected at a rate
$\lambda = \beta \nu \TI$ $yrs^{-1}$, where $\beta$ is the transmission coefficient
and $\nu$ is the proportion of pulmonary TB cases. A proportion $\phi$ and
$(1-\phi)$ of individuals in the class $\TP$ is transferred to the class $\TI$
and $\TL$, respectively, at a rate $\delta \, yrs^{-1}$. Each year, a proportion
$k$ of individuals in the class $\TI$ is detected and start TB treatment
at a rate $\tau \, yrs^{-1}$, entering the class $\TT$. It is assumed that
individuals in the class $\TT$ are neither infectious nor susceptible to reinfection.
A fraction $\phi_T$ of individuals in class $\TT$ is transferred to class $\TI$
due to either treatment failure or default, while the remaining $(1-\phi_T)$
are successfully treated and enter in the class $\TL$. The inverse of treatment
length is denoted by $\delta_T$. In \cite{TBportugalGomesRodrigues}, birth and
death rates are assumed equal, here we assume that they can be different and we
denote the recruitment rate by $\eta \, yrs^{-1}$ and the death rate by
$\mu \, yrs^{-1}$. The reinfection factor is denoted by $\sigma$
(see \cite{TBportugalGomesRodrigues} for more details). Optimal control strategies
for such model were studied in \cite{MR3266821,SilvaTorresTBAngola,MR3101449}.

Let $\TS\equiv \TS(t)$, $\TP\equiv \TP(t)$, $\TI\equiv \TI(t)$,
$\TL\equiv \TL(t)$, $\TT\equiv \TT(t)$, where $t$ represents time in years.
The model described above is given by the following system
of ordinary differential equations:
\myeq{eq1}{
\left\{\begin{array}{l}
\dot{\TS}= \eta N - \left(\lambda(t)+\mu\right)\TS,\\
\dot{\TP}=\lambda(t)\TS+\sigma\lambda(t)\TL-\left(\delta+\mu\right)\TP,\\
\dot{\TI}=\phi\delta \TP+\omega \TL+\phi_T\delta_T \TT-\left(\tau k + \mu\right)\TI,\\
\dot{\TL}=(1-\phi)\delta \TP+(1-\phi_T)\delta_T \TT
-\left(\sigma\lambda(t)+\omega+\mu\right)\TL,\\
\dot{\TT}=\tau k \TI-\left(\delta_T+\mu\right)\TT.
\end{array}\right.
}
We have $N=\TS+\TP+\TI+\TL+\TT$ and $\lambda(t)=\beta\nu \TI N^{-1}$. Then
$$
\dot{\lambda}=\beta\nu\left(\dot{\TI}N^{-1}-\TI\,N^{-2}\dot{N}\right).
$$
On the other hand, $\dot{N}=(\eta-\mu)N$, so if $\eta=\mu$ then the population
is constant. The system can be written in a matrix form as
\myeq{eq1a}{
\dot{\TX}=\left(\beta\nu \TI \mathcal{A}+\mathcal{B}\right)\TX+\mathcal{C},
}
where $\TX=(\TS,\TP,\TI,\TL,\TT)$,
$$  \mathcal{A}=\left(\begin{array}{ccccc}
-1 & 0 & 0 & 0 & 0\\
1 & 0 & 0 & \sigma & 0\\
0 & 0 & 0 & 0 & 0\\
0 & 0 & 0 & -\sigma & 0\\
0 & 0 & 0 & 0 & 0
\end{array}\right),
\quad
 \mathcal{B}=\left(\begin{array}{ccccc}
-\mu & 0 & 0 & 0 & 0\\
0 & -(\delta+\mu) & 0 & 0 & 0\\
0 & \phi\delta & -(\tau k + \mu) & \omega & \phi_T\delta_T\\
0 & (1-\phi)\delta & 0 & -(\omega+\mu) & (1-\phi_T)\delta_T\\
0 & 0 & \tau k & 0 & -(\delta_T+\mu)
\end{array}\right),
$$
and $\mathcal{C}=(\eta N, 0, 0, 0, 0)$. We can verify that the matrix
$\lambda(t)\mathcal{A}+\mathcal{B}$ can be diagonalizable, so there is a
semi-closed form solution for the problem (it is not closed a priori because
$\lambda$ still depends on $I$ and $N$).

Suppose this system interacts with (a convex combination of) another two similar
systems $\tilde{X}_1$ and $\tilde{X}_2$, in the following way: there exist
functions $\gamma(t), \tilde{\gamma}(t)\in[0,1]$
and a value $\zeta\in[0,1]$ such that
\myeq{eq2}{
\left\{\begin{array}{l}
\dot{\TS}= \eta N - \left(\lambda(t)+\gamma(t)+\mu\right)\TS
+\tilde{\gamma}(t)\left((1-\zeta)\tilde{\TS}_1+\zeta\tilde{\TS}_2\right),\\
\dot{\TP}=\lambda(t)\TS+\sigma\lambda(t)\TL-\left(\delta+\gamma(t)+\mu\right)\TP
+\tilde{\gamma}(t)\left((1-\zeta)\tilde{\TP}_1+\zeta\tilde{\TP}_2\right),\\
\dot{\TI}=\phi\delta \TP+\omega \TL+\phi_T\delta_T \TT-\left(\tau k + \gamma(t)
+\mu\right)\TI+\tilde{\gamma}(t)\left((1-\zeta)\tilde{\TI}_1+\zeta\tilde{\TI}_2\right),\\
\dot{\TL}=(1-\phi)\delta \TP+(1-\phi_T)\delta_T \TT-\left(\sigma\lambda(t)
+\omega+\gamma(t)+\mu\right)\TL+\tilde{\gamma}(t)\left((1-\zeta)\tilde{\TL}_1
+\zeta\tilde{\TL}_2\right),\\
\dot{\TT}=\tau k \TI-\left(\delta_T+\gamma(t)+\mu\right)\TT
+\tilde{\gamma}(t)\left((1-\zeta)\tilde{\TT}_1+\zeta\tilde{\TT}_2\right).
\end{array}\right.
}
Adding $N=\TS+\TP+\TI+\TL+\TT$ as a new state variable, we have
\myeq{eq3}{
\left\{\begin{array}{l}
\dot{\TS}= \eta N - \left(\lambda +\gamma(t)+\mu\right)\TS
+\tilde{\gamma}(t)\left((1-\zeta)\tilde{\TS}_1+\zeta\tilde{\TS}_2\right),\\
\dot{\TP}=\lambda \TS+\sigma\lambda \TL-\left(\delta+\gamma(t)+\mu\right)\TP
+\tilde{\gamma}(t)\left((1-\zeta)\tilde{\TP}_1+\zeta\tilde{\TP}_2\right),\\
\dot{\TI}=\phi\delta \TP+\omega \TL+\phi_T\delta_T \TT-\left(\tau k + \gamma(t)
+\mu\right)\TI+\tilde{\gamma}(t)\left((1-\zeta)\tilde{\TI}_1+\zeta\tilde{\TI}_2\right),\\
\dot{\TL}=(1-\phi)\delta \TP+(1-\phi_T)\delta_T \TT-\left(\sigma\lambda+\omega
+\gamma(t)+\mu\right)\TL+\tilde{\gamma}(t)\left((1-\zeta)\tilde{\TL}_1
+\zeta\tilde{\TL}_2\right),\\
\dot{\TT}=\tau k \TI-\left(\delta_T+\gamma(t)+\mu\right)\TT
+\tilde{\gamma}(t)\left((1-\zeta)\tilde{\TT}_1+\zeta\tilde{\TT}_2\right),\\
\dot{N}=\left(\eta-\gamma(t)-\mu\right)N
+\tilde{\gamma}(t)\left((1-\zeta)\tilde{N}_1+\zeta\tilde{N}_2\right).
\end{array}\right.
}
Let $S=\TS N^{-1}$, $P=\TP N^{-1}$, $I=\TI N^{-1}$, $L=\TL N^{-1}$,
$T=\TT N^{-1}$. These variables now represent the percentage of the
population in each state, i.e., $S+P+I+L+T=1$. Since
\myeqAN{
\dot{S}&=\dot{\TS}N^{-1}-\TS N^{-2}\dot{N}\\
&=\dot{\TS}N^{-1}-S N^{-1}\left(\left(\eta-\gamma(t)
-\mu\right)N+\tilde{\gamma}(t)\left((1-\zeta)\tilde{N}_1
+\zeta\tilde{N}_2\right)\right)\\
&=\dot{\TS}N^{-1}-\left(\eta-\gamma(t)-\mu+\tilde{\gamma}(t)\left((1-\zeta)
\tilde{N}_1+\zeta\tilde{N}_2\right){N}^{-1}\right)S,\\
&=\dot{\TS}N^{-1}-\left(M(t)-\gamma(t)-\mu\right)S,
}
with $M(t)\mdef\eta+\left((1-\zeta)\tilde{N}_1
+\zeta\tilde{N}_2\right)\tilde{\gamma}(t)N^{-1}$,
where the calculations for the other variables are similar, and adding
$\lambda(t)=\beta\nu I$ as a new state variable, we have
\myeq{eq4}{
\left\{\begin{array}{l}
\dot{S}= \eta - \left(\lambda+M(t)\right)S
+\tilde{\gamma}(t)\left((1-\zeta)\tilde{S}_1+\zeta\tilde{S}_2\right),\\
\dot{P}=\lambda S+\sigma\lambda L-\left(\delta+M(t)\right)P
+\tilde{\gamma}(t)\left((1-\zeta)\tilde{P}_1+\zeta\tilde{P}_2\right),\\
\dot{I}=\phi\delta P+\omega L+\phi_T\delta_T T-\left(\tau k +M(t)\right)I
+\tilde{\gamma}(t)\left((1-\zeta)\tilde{I}_1+\zeta\tilde{I}_2\right),\\
\dot{L}=(1-\phi)\delta P+(1-\phi_T)\delta_T T-\left(\sigma\lambda+\omega
+M(t)\right)L+\tilde{\gamma}(t)\left((1-\zeta)\tilde{L}_1+\zeta\tilde{L}_2\right),\\
\dot{T}=\tau k I-\left(\delta_T+M(t)\right)T
+\tilde{\gamma}(t)\left((1-\zeta)\tilde{T}_1+\zeta\tilde{T}_2\right),\\
\dot{\lambda}=\beta\nu\dot{I}=\beta\nu\left(\phi\delta P+\omega L
+\phi_T\delta_T T-\left(\tau k+M(t)\right)I
+\tilde{\gamma}(t)\left((1-\zeta)\tilde{I}_1+\zeta\tilde{I}_2\right)\right),\\
\dot{N}=\left(M(t)-\gamma(t)-\mu\right)N.
\end{array}\right.
}
Using the above model, we consider different population groups: people living
in a high incidence TB country (A) and people living in a low incidence TB
country (B), where (B) is subdivided in a community (G) with high percentage
of people from (A), and (C) is the rest of the population of (B).
We consider that the values of $\beta$, $\nu$, $\phi_T$ of the group (G) are
different from the values of the group (C). The flux of population follow
the distribution functions $\gamma_{A}$, from (A) to (B), and $\gamma_{B}$,
from (B) to (A). We assume that the persons that arrive and departure from (B)
are split in the following proportions: $\zeta$ goes to (G) and $(1-\zeta)$ goes
to (C), with $\zeta\in[0,1]$ a fixed percentage value in this model.

This model accounts for an average moving value of persons $a^A$, $a^B$ that
increases/decreases in time by the slopes $b^A$, $b^B$ and has a seasonality
variation modeled by $p^A$, $p^B$, $\theta^A$, $\theta^B$. The flux
of population will be modeled by the following functions:
\myeq{eq5a}{
\gamma_A(t) = a^A + b^A t +  a^A p^A \cos\left(\theta^A t\right),
\quad \mbox{ and } \quad
\gamma_B(t) = a^B + b^B t +  a^B p^B \cos\left(\theta^B t\right),
}
for constants $a^A, a^B, b^A, b^B, p^A, p^B, \theta^A, \theta^B\in\bkR$
chosen to ensure that $0\leq \gamma_A(t),\gamma_B(t)\leq 1$ for all $t$
of the simulation.

The flux of population $\gamma_A(t)$, $\gamma_B(t)$ can be incorporated
as state-space variables. In our case, the functions $\gamma^A$, $\gamma^P$
are solutions of the system of ODEs
$$
\left\{\begin{array}{l}
\dot{\gamma}_A = z^A,\\
\dot{z}_A = -(\theta^A)^2(\gamma_A- a^A- b^A\,t),
\end{array}\right.
\quad \mbox{ and } \quad
\left\{\begin{array}{l}
\dot{\gamma}_B = z^B,\\
\dot{z}_B = -(\theta^B)^2(x- a^B- b^B\,t),
\end{array}\right.
$$
which we add to the model~\myref{eqPA}--\myref{eqPGamma}, obtaining the complete model
with $25$ state-space variables. Note that if $V_N = (N_A,N_C,N_G)$, then
\myeq{eq6}{
\dot{V}_N=\mathcal{A}(t) V_N,
}
where
$$
\mathcal{A}(t)=\left(\begin{array}{ccc}
\eta^A-\mu^A-\gamma_A(t) & \gamma_B(t)(1-\zeta) & \gamma_B(t)\zeta\\
\gamma_A(t) (1-\zeta) & \eta^C-\mu^C-\gamma_B(t) & 0\\
\gamma_A(t) \zeta & 0 & \eta^C-\mu^C-\gamma_B(t)
\end{array}\right).
$$
So the population evolution is only dependent on the moving distribution
functions $\gamma^A$, $\gamma^P$, born rates $\eta$, and natural death rates $\mu$.
Hence, we obtain the complete model composed by the four subsystems 
\myref{eqPA}--\myref{eqPGamma} composed by: (i) the variables
of the high incidence TB country
\myeq{eqPA}{
\left\{\begin{array}{l}
\dot{S}_A= \eta^A - \left(\lambda_A +M_A\right)S_A+\gamma_B\left((1-\zeta)
S_C+\zeta S_G\right),\\
\dot{P}_A=\lambda_A S_A+\sigma^A\lambda_A L_A-\left(\delta^A+M_A\right)P_A
+\gamma_B\left((1-\zeta) P_C+\zeta P_G\right),\\
\dot{I}_A=\phi^A\delta^A P_A+\omega^A L_A+\phi^A_T\delta^A_T T_A
-\left(\tau^A k^A +M_A\right)I_A+\gamma_B\left((1-\zeta) I_C
+\zeta I_G\right),\\
\dot{L}_A=(1-\phi^A)\delta^A P_A+(1-\phi^A_T)\delta^A_T T_A
-\left(\sigma^A\lambda_A+\omega^A+M_A\right)L_A+\gamma_B\left((1-\zeta) L_C
+\zeta L_G\right),\\
\dot{T}_A=\tau^A k^A I_A-\left(\delta^A_T+M_A\right)T_A
+\gamma_B\left((1-\zeta) T_C+\zeta T_G\right),\\
\dot{\lambda}_A=\beta^A\nu^A\left(\phi^A\delta^A P_A+\omega^A L_A
+\phi^A_T\delta^A_T T_A-\left(\tau^A k^A+M_A\right)I^A
+\gamma_B\left((1-\zeta) I_C+\zeta I_G\right)\right),\\
\dot{N}_A=\left(M_A-\gamma_A-\mu^A\right)N_A,\\
\end{array}\right.
}
(ii) the variables associated with the community in the host country
\myeq{eqPG}{
\left\{\begin{array}{l}
\dot{S}_G= \eta^C - \left(\lambda_G +M_G\right)S_G+\gamma_A\zeta S_A,\\
\dot{P}_G=\lambda_G S_G+\sigma^C\lambda_G L_G-\left(\delta^C+M_G\right)P_G
+\gamma_A\zeta P_A,\\
\dot{I}_G=\phi^C\delta^C P_G+\omega^C L_G+\phi^G_T\delta^C_T T_G
-\left(\tau^C k^C +M_G\right)I_G+\gamma_A\zeta I_A,\\
\dot{L}_G=(1-\phi^C)\delta^C P_G+(1-\phi^G_T)\delta^C_T T_G
-\left(\sigma^C\lambda_G+\omega^C+M_G\right)L_G+\gamma_A\zeta L_A,\\
\dot{T}_G=\tau^C k^C I_G-\left(\delta^C_T+M_G\right)T_G+\gamma_A\zeta T_A,\\
\dot{\lambda}_G=\beta^G\nu^G\left(\phi^C\delta^C P_G+\omega^C L_G
+\phi^G_T\delta^C_T T_G-\left(\tau^C k^C+M_G\right)I^G+\gamma_A\zeta I_A\right),\\
\dot{N}_G=\left(M_G-\gamma_B-\mu^C\right)N_G,\\
\end{array}\right.
}
(iii) the variables related with the population of the host country excluding
the community 
\myeq{eqPC}{
\left\{\begin{array}{l}
\dot{S}_C= \eta^C - \left(\lambda_C +M_C\right)S_C+\gamma_A(1-\zeta) S_A,\\
\dot{P}_C=\lambda_C S_C+\sigma^C\lambda_C L_C-\left(\delta^C+M_C\right)P_C
+\gamma_A(1-\zeta) P_A,\\
\dot{I}_C=\phi^C\delta^C P_C+\omega^C L_C+\phi^C_T\delta^C_T T_C
-\left(\tau^C k^C +M_C\right)I_C+\gamma_A(1-\zeta) I_A,\\
\dot{L}_C=(1-\phi^C)\delta^C P_C+(1-\phi^C_T)\delta^C_T T_C
-\left(\sigma^C\lambda_C+\omega^C+M_C\right)L_C+\gamma_A(1-\zeta) L_A,\\
\dot{T}_C=\tau^C k^C I_C-\left(\delta^C_T+M_C\right)T_C+\gamma_A(1-\zeta) T_A,\\
\dot{\lambda}_C=\beta^C\nu^C\left(\phi^C\delta^C P_C+\omega^C L_C
+\phi^C_T\delta^C_T T_C-\left(\tau^C k^C+M_C\right)I^C+\gamma_A(1-\zeta) I_A\right),\\
\dot{N}_C=\left(M_C-\gamma_B-\mu^C\right)N_C,\\
\end{array}\right.
}
(iv) and the variables measuring the flux of population
\myeq{eqPGamma}{
\left\{\begin{array}{l}
\dot{\gamma}_A = z^A,\\
\dot{z}_A = -(\theta^A)^2(\gamma_A- a^A-b^A\,t),\\
\dot{\gamma}_B = z^B,\\
\dot{z}_B = -(\theta^B)^2(\gamma_B- a^B- b^B\,t),
\end{array}\right.
}
where for presentation convenience we define
\myeqAN{
M_A&=\eta^A+\left((1-\zeta) N_C+\zeta N_G\right)\gamma_B N_A^{-1},\\
M_C&=\eta^C+(1-\zeta) \gamma_A N_A N_C^{-1},\\
M_G&=\eta^C+\zeta \gamma_A N_A N_G^{-1}.
}
Note that
\myeqAN{
\dot{N}_A+\dot{N}_C+\dot{N}_G &=(\eta^A-\mu^A)N_A+(\eta^C-\mu^C)(N_C+N_G).
}
Again, if $\eta^A=\mu^A$ and $\eta^C=\mu^C$, then the total population is constant.
Moreover, if $b^A=b^B=p^A=p^B= 0$, then system \myref{eqPA}--\myref{eqPGamma} is autonomous. For
notation clarity, all parameters (i.e., constant values) have upper indices
whereas state variables have lower indices.

\begin{figure}
\centering
\scalebox{0.90}
{
\begin{pspicture}(0,-3.1489062)(8.3828125,3.1289062)
\psframe[linewidth=0.04,dimen=outer](2.8609376,2.3489063)(1.2809376,1.5089062)
\psframe[linewidth=0.04,dimen=outer](2.8609376,0.34890625)(1.2809376,-0.49109375)
\psframe[linewidth=0.04,dimen=outer](6.6609373,0.32890624)(5.0809374,-0.51109374)
\psframe[linewidth=0.04,dimen=outer](6.6809373,-2.0310938)(5.1009374,-2.8710938)
\psframe[linewidth=0.04,dimen=outer](2.8809376,-2.0310938)(1.3009375,-2.8710938)
\psline[linewidth=0.04cm,arrowsize=0.05291667cm 2.0,arrowlength=1.4,
arrowinset=0.4]{->}(2.0809374,1.4489063)(2.0809374,0.42890626)
\psline[linewidth=0.04cm,arrowsize=0.05291667cm 2.0,arrowlength=1.4,
arrowinset=0.4]{->}(2.0609374,-0.55109376)(2.0809374,-2.0510938)
\psline[linewidth=0.04cm,arrowsize=0.05291667cm 2.0,arrowlength=1.4,
arrowinset=0.4]{<-}(5.8809376,-0.55109376)(5.9009376,-2.0510938)
\psline[linewidth=0.04cm,arrowsize=0.05291667cm 2.0,arrowlength=1.4,
arrowinset=0.4]{->}(2.9609375,0.14890625)(5.0409374,0.14890625)
\psline[linewidth=0.04cm,arrowsize=0.05291667cm 2.0,arrowlength=1.4,
arrowinset=0.4]{<-}(2.9609375,-0.27109376)(5.0409374,-0.27109376)
\psline[linewidth=0.04cm,arrowsize=0.05291667cm 2.0,arrowlength=1.4,
arrowinset=0.4]{<-}(2.9409375,-2.6710937)(5.0209374,-2.6710937)
\psline[linewidth=0.04cm,arrowsize=0.05291667cm 2.0,arrowlength=1.4,
arrowinset=0.4]{->}(2.9809375,-2.2710938)(5.0609374,-2.2710938)
\psline[linewidth=0.04cm,arrowsize=0.05291667cm 2.0,arrowlength=1.4,
arrowinset=0.4]{->}(2.0809374,3.1089063)(2.1009376,2.3689063)
\psline[linewidth=0.04cm,arrowsize=0.05291667cm 2.0,arrowlength=1.4,
arrowinset=0.4]{->}(5.3409376,-0.55109376)(2.3809376,-1.9710938)
\usefont{T1}{ptm}{m}{n}
\rput(2.1023438,1.9189062){$S$}
\usefont{T1}{ptm}{m}{n}
\rput(2.0723438,-0.06109375){$P$}
\usefont{T1}{ptm}{m}{n}
\rput(5.8923435,-0.10109375){$L$}
\usefont{T1}{ptm}{m}{n}
\rput(2.0223436,-2.4210937){$I$}
\usefont{T1}{ptm}{m}{n}
\rput(5.882344,-2.4410937){$T$}
\psline[linewidth=0.04cm,arrowsize=0.05291667cm 2.0,arrowlength=1.4,
arrowinset=0.4]{->}(1.1809375,2.1489062)(0.4409375,2.1489062)
\psline[linewidth=0.04cm,arrowsize=0.05291667cm 2.0,arrowlength=1.4,
arrowinset=0.4]{->}(1.2209375,0.12890625)(0.4809375,0.12890625)
\psline[linewidth=0.04cm,arrowsize=0.05291667cm 2.0,arrowlength=1.4,
arrowinset=0.4]{->}(1.2409375,-2.2710938)(0.5009375,-2.2710938)
\psline[linewidth=0.04cm,arrowsize=0.05291667cm 2.0,arrowlength=1.4,
arrowinset=0.4]{<-}(1.2209375,1.9089062)(0.4809375,1.9089062)
\psline[linewidth=0.04cm,arrowsize=0.05291667cm 2.0,arrowlength=1.4,
arrowinset=0.4]{<-}(1.2609375,-0.09109375)(0.5209375,-0.09109375)
\psline[linewidth=0.04cm,arrowsize=0.05291667cm 2.0,arrowlength=1.4,
arrowinset=0.4]{<-}(1.2809376,-2.4710937)(0.5409375,-2.4710937)
\psline[linewidth=0.04cm,arrowsize=0.05291667cm 2.0,arrowlength=1.4,
arrowinset=0.4]{<-}(7.5209374,0.14890625)(6.7809377,0.14890625)
\psline[linewidth=0.04cm,arrowsize=0.05291667cm 2.0,arrowlength=1.4,
arrowinset=0.4]{<-}(7.5009375,-2.2510939)(6.7609377,-2.2510939)
\psline[linewidth=0.04cm,arrowsize=0.05291667cm 2.0,arrowlength=1.4,
arrowinset=0.4]{->}(7.5209374,-2.4910936)(6.7809377,-2.4910936)
\psline[linewidth=0.04cm,arrowsize=0.05291667cm 2.0,arrowlength=1.4,
arrowinset=0.4]{->}(7.5209374,-0.07109375)(6.7809377,-0.07109375)
\usefont{T1}{ptm}{m}{n}
\rput(0.85234374,2.3989062){$M$}
\usefont{T1}{ptm}{m}{n}
\rput(0.83234376,0.33890626){$M$}
\usefont{T1}{ptm}{m}{n}
\rput(0.8723438,-2.0410938){$M$}
\usefont{T1}{ptm}{m}{n}
\rput(7.2323437,-1.9810938){$M$}
\usefont{T1}{ptm}{m}{n}
\rput(7.2323437,0.37890625){$M$}
\usefont{T1}{ptm}{m}{n}
\rput(0.8123438,1.6789062){$\tilde{S}$}
\usefont{T1}{ptm}{m}{n}
\rput(0.78234375,-0.34109375){$\tilde{P}$}
\usefont{T1}{ptm}{m}{n}
\rput(0.73234373,-2.7210937){$\tilde{I}$}
\usefont{T1}{ptm}{m}{n}
\rput(7.202344,-0.32109374){$\tilde{L}$}
\usefont{T1}{ptm}{m}{n}
\rput(7.2123437,-2.7210937){$\tilde{T}$}
\usefont{T1}{ptm}{m}{n}
\rput(2.4623437,0.93890625){$\lambda$}
\usefont{T1}{ptm}{m}{n}
\rput(4.092344,-0.48109376){$\sigma \lambda$}
\usefont{T1}{ptm}{m}{n}
\rput(3.9523437,0.37890625){$(1-\phi)\delta$}
\usefont{T1}{ptm}{m}{n}
\rput(6.8123436,-1.2810937){$(1-\phi_T)\delta_T$}
\usefont{T1}{ptm}{m}{n}
\rput(1.7923437,-1.1610937){$\phi \delta$}
\usefont{T1}{ptm}{m}{n}
\rput(4.032344,-2.9210937){$\phi_T \delta_T$}
\usefont{T1}{ptm}{m}{n}
\rput(3.9923437,-2.0610938){$\tau k$}
\usefont{T1}{ptm}{m}{n}
\rput{25.0}(-0.101552695,-1.7434878){\rput(3.8623438,-1.0810938){$\omega$}}
\usefont{T1}{ptm}{m}{n}
\rput(2.3223438,2.7389061){$\eta$}
\end{pspicture}
}
\caption{Model for TB transmission.}
\label{fig:model:flow1}
\end{figure}
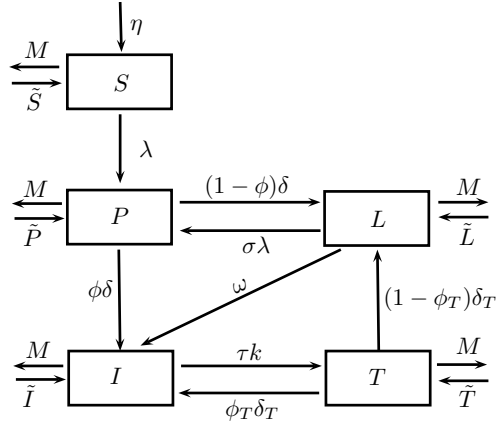

\begin{figure}
\centering
\scalebox{0.90}
{
\begin{pspicture}(0,-1.39)(10.063281,1.39)
\psframe[linewidth=0.04,dimen=outer](9.82,0.79)(8.22,-0.81)
\usefont{T1}{ptm}{m}{n}
\rput(4.5746875,0.005){\large $A$}
\usefont{T1}{ptm}{m}{n}
\rput(0.79468745,-0.035){\large $G$}
\usefont{T1}{ptm}{m}{n}
\rput(9.024688,0.045){\large $C$}
\psline[linewidth=0.04cm,arrowsize=0.05291667cm 2.0,arrowlength=1.4,
arrowinset=0.4]{->}(4.64,1.37)(4.64,0.87)
\usefont{T1}{ptm}{m}{n}
\rput(4.9028125,1.2){$\eta^A$}
\psline[linewidth=0.04cm,arrowsize=0.05291667cm 2.0,arrowlength=1.4,
arrowinset=0.4]{<-}(3.76,0.23)(1.74,0.23)
\psline[linewidth=0.04cm,arrowsize=0.05291667cm 2.0,arrowlength=1.4,
arrowinset=0.4]{<-}(8.12,0.21)(5.58,0.21)
\psline[linewidth=0.04cm,arrowsize=0.05291667cm 2.0,arrowlength=1.4,
arrowinset=0.4]{->}(8.0,-0.21)(5.54,-0.21)
\psline[linewidth=0.04cm,arrowsize=0.05291667cm 2.0,arrowlength=1.4,
arrowinset=0.4]{->}(3.74,-0.21)(1.72,-0.21)
\psline[linewidth=0.04cm,arrowsize=0.05291667cm 2.0,arrowlength=1.4,
arrowinset=0.4]{->}(0.92,1.35)(0.92,0.85)
\psline[linewidth=0.04cm,arrowsize=0.05291667cm 2.0,arrowlength=1.4,
arrowinset=0.4]{->}(9.08,1.29)(9.08,0.79)
\usefont{T1}{ptm}{m}{n}
\rput(9.3828125,1.16){$\eta^C$}
\usefont{T1}{ptm}{m}{n}
\rput(1.2128125,1.2){$\eta^G$}
\usefont{T1}{ptm}{m}{n}
\rput(2.7728124,-0.46){$\zeta \gamma_A X_A$}
\psline[linewidth=0.04cm,arrowsize=0.05291667cm 2.0,arrowlength=1.4,
arrowinset=0.4]{->}(0.86,-0.83)(0.86,-1.33)
\psline[linewidth=0.04cm,arrowsize=0.05291667cm 2.0,arrowlength=1.4,
arrowinset=0.4]{->}(4.64,-0.87)(4.64,-1.37)
\psline[linewidth=0.04cm,arrowsize=0.05291667cm 2.0,arrowlength=1.4,
arrowinset=0.4]{->}(9.04,-0.85)(9.04,-1.35)
\usefont{T1}{ptm}{m}{n}
\rput(1.2428125,-1.04){$M_G$}
\usefont{T1}{ptm}{m}{n}
\rput(5.0128126,-1.08){$M_A$}
\usefont{T1}{ptm}{m}{n}
\rput(9.452812,-1.04){$M_C$}
\usefont{T1}{ptm}{m}{n}
\rput(2.7728124,0.46){$\zeta \gamma_B X_G$}
\usefont{T1}{ptm}{m}{n}
\rput(6.732813,0.46){$(1-\zeta)\gamma_A X_A$}
\usefont{T1}{ptm}{m}{n}
\rput(6.742812,-0.44){$(1-\zeta)\gamma_B X_C$}
\psframe[linewidth=0.04,dimen=outer](5.42,0.79)(3.82,-0.81)
\psframe[linewidth=0.04,dimen=outer](1.6,0.77)(0.0,-0.83)
\end{pspicture}
}
\caption{Flow chart between high TB incidence country (A), natives from high
TB incidence country living in Communities (G) in a low TB incidence country,
remainder of population living in a low TB incidence country (C).}
\label{fig:model:flow2}
\end{figure}
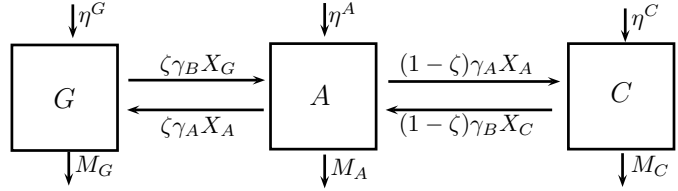

% ------------------------------------------

\section{Reproduction number and its sensitivity analysis for the autonomous case}
\label{sec:R0}

The transmissibility of an infection can be asymptotically quantified by its
reproduction number $R_0$ (for autonomous models), defined as the mean number
of secondary infections seeded by a typical infective into a susceptible
population. Since $R_0$ is a condition for the asymptotic stability of solutions
around a free disease equilibrium point, this value determines a threshold:
whenever $R_0 > 1$, a typical infective gives rise, on average, to more than
one secondary infection, leading to an epidemic. In contrast, when $R_0 < 1$,
infectious typically give rise, on average, to less than one secondary infection,
and the prevalence of infection cannot increase.

A key point is that the model \myref{eqPA}--\myref{eqPGamma} is \emph{a priori} nonautonomous,
due to the flux of population $\gamma_A$ and $\gamma_B$. For such reason,
from now on we assume that $\gamma_A(t)\equiv a^A$ and $\gamma_B(t)=a^B$,
i.e., $b^A=b^B=p^A=p^B=0$ in~\myref{eq5a}, so that model \myref{eqPA}--\myref{eqPGamma} becomes
autonomous and we can apply the standard method from \cite{Driessche-Watmough}.
A complete nonautonomous situation will be considered in a future work.

The reproducing number $R_0$ of system~\myref{eq1} can be analytically
determined and, when $\eta=\mu$, is given by
\myeq{eqR0s}{
R_0 = \frac{\beta\nu\delta(\delta_T+\mu)(\phi\mu+\omega)}{\mu(\delta+\mu)[
(\mu+\omega)(\tau k+\delta_T+\mu)+\delta_T \tau k(1-\phi_T)]},
}
see, e.g., \cite{TBportugalGomesRodrigues}. Hence, $R_0$ is proportional to
$\beta$, $\nu$, $\phi$, $\phi_T$ ($0<\phi_T<1$) and inverse proportional to
$\tau$ and $k$. In the no-transfer situation, i.e., $\gamma_A\equiv \gamma_B
\equiv 0$, our model reduces to the disjoint coupling of the (sub)systems
$(A)$, $(C)$ and $(G)$ similar to~\myref{eq1}, so we can compute the
reproduction numbers for the subsystems (using the fixed parameters
from Table~\ref{table:parameters:PT:AN}) in the no-transfer situation
using~\myref{eqR0s}, which gives
$$
R_0^A=\numD{6.784924946},\quad R_0^C=\numD{1.116995163},
\quad R_0^G=\numD{2.365451295},
$$
where $R_0^A$, $R_0^C$ and $R_0^G$ denote the basic reproduction number for
populations (A), (C) and (G), respectively, when they are complete independent
from each others (no flux of population between the compartments). For the
complete system \myref{eqPA}--\myref{eqPGamma} the basic reproduction number will be denoted
by $R_0^T$. Note that the coupling of only $(C)$ and $(G)$ (again in the
no-transfer situation and without the components associated to $(A)$) is known
in the literature as a model for heterogeneous infection risk
\cite{Gomes:etall:2012,TBportugalGomesRodrigues}.

The complete system~\myref{eqPA}--\myref{eqPGamma}, although a generalization of previous models,
is quite different from systems like~\myref{eq1}, by the fact that it has
internal transfer of individuals between subsystems $(A)$ and $(C)$ and $(G)$,
so it is not expected that $R_0^T$ follows the same expression~\myref{eqR0s}.
So its relevant to understand how $R_0^T$ is affected by variation of the
parameters. In order to verify the validity and to obtain the value of $R_0^T$,
depending on the parameters chosen, we follow the approach in~\cite{Driessche-Watmough}.

Let $x$ represent the state-space variables (in a special order) that group the
individuals in each disease state and group compartment, i.e.,
$$
x=(P_A,P_C,P_G,I_A,I_C,I_G,L_A,L_C,L_G,T_A,T_C,T_G,S_A,S_C,S_G)\in\bkR^{15}_+.
$$
Note that there exists an equilibrium point with $I_A,I_C,I_G=0$,
if $\lambda_A=\lambda_B=\lambda_C=0$ and
$$
\left\{\begin{array}{l}
\eta^A- M_A S_A+a^B\left((1-\zeta) S_C+\zeta S_G\right)=0,\\
\eta^C- M_C S_C+a^A(1-\zeta) S_A=0,\\
\eta^G- M_G S_G+a^A\zeta S_A=0,\\
-\left(\delta^A+M_A\right)P_A+a^B\left((1-\zeta) P_C+\zeta P_G\right)=0,\\
-\left(\delta^C+M_C\right)P_C+a^A(1-\zeta) P_A=0,\\
-\left(\delta^C+M_G\right)P_G+a^A\zeta P_A=0,\\
\phi^A\delta^A P_A+\omega^A L_A+\phi^A_T\delta^A_T T_A=0,\\
\phi^C\delta^C P_C+\omega^C L_C+\phi^C_T\delta^C_T T_C=0,\\
\phi^C\delta^C P_G+\omega^C L_G+\phi^G_T\delta^C_T T_G=0,\\
(1-\phi^A)\delta^A P_A+(1-\phi^A_T)\delta^A_T T_A
-\left(\omega^A+M_A\right)L_A+a^B\left((1-\zeta) L_C
+\zeta L_G\right)=0,\\
(1-\phi^C)\delta^C P_C+(1-\phi^C_T)\delta^C_T T_C
-\left(\omega^C+M_C\right)L_C+a^A(1-\zeta) L_A=0,\\
(1-\phi^C)\delta^C P_G+(1-\phi^G_T)\delta^C_T T_G
-\left(\omega^C+M_G\right)L_G+a^A\zeta L_A=0,\\
-\left(\delta^A_T+M_A\right)T_A+a^B\left((1-\zeta) T_C+\zeta T_G\right)=0,\\
-\left(\delta^C_T+M_C\right)T_C+a^A(1-\zeta) T_A=0,\\
-\left(\delta^C_T+M_G\right)T_G+a^A\zeta T_A=0.
\end{array}\right.
$$
From the last three equations, we have
$$ \left(\begin{array}{ccc}
-\delta^A_T-M_A & a^B(1-\zeta) & a^B\zeta\\
a^A(1-\zeta) & -\delta^C_T-M_C & 0\\
a^A\zeta & 0 & -\delta^G_T-M_G
\end{array}\right)
\left(\begin{array}{c}
T_A\\ T_C\\T_G
\end{array}\right)
=
\left(\begin{array}{c}
0 \\ 0 \\ 0
\end{array}\right)
\:\:\Rightarrow\:\:T_A=T_C=T_G=0.
$$
In the same way we can see, from fourth to sixth equations,
that $P_A=P_C=P_G=0$ and, from the other equations,
that $L_A=L_C=L_G=0$. Since, $\eta^A, \eta^C\neq0$ and
\myeqAN{
M_A S_A&=\eta^A S_A+\left((1-\zeta) S_C+\zeta S_G\right)a^B,\\
M_C S_C&=\eta^C S_C+(1-\zeta) a^A S_A,\\
M_G S_G&=\eta^C S_G +\zeta a^A S_A,
}
from the first three equations, we have $S_A= S_C = S_G = 1$. Hence,
the disease free equilibrium point (DFE) is unique and given by
$$
x_0=(0,0,0,0,0,0,0,0,0,0,0,0,1,1,1),
$$
and it makes sense to define the set of all disease free states $X_s$ as
$$
X_s=\{(0,0,0,0,0,0,0,0,0,0,0,0,S_A,S_C,S_G)\in\bkR^{15}\::\: S_A, S_C, S_G\geq 0\}.
$$

In our model the individuals get the first contact with the infection in the
states $P_A, P_C, P_G$. We have $m=12$ states where individuals have different
degrees of infection and $3$ states free of disease. The vector field $X$
in~\myref{eqPA}--\myref{eqPGamma} is now divided as $X=\mathcal{F}-(\mathcal{V}^--\mathcal{V}^+)$,
where $\mathcal{F}$ is the rate of appearance of new infections, $\mathcal{V}^+$
is the rate of in-transfers of individuals by other means, and $\mathcal{V}^-$
is the rate of out-transfers of individuals by other means. We have
{\footnotesize
$$
\mathcal{F}_{1-3}(x)
=\left(\begin{array}{c}
\beta^A\nu^A I_A \left(S_A+\sigma^A L_A\right)
+a^B \left((1-\zeta) P_C+\zeta P_G\right)\\
\beta^C\nu^C I_C \left(S_C+\sigma^C L_C\right)+a^A(1-\zeta) P_A\\
\beta^G\nu^G I_G \left(S_G+\sigma^C L_G\right)+a^A\zeta P_A
\end{array}\right), \quad \mathcal{F}_j(x) = 0
\mbox { for  } j\in\{4,\cdots, 15\},
$$
$$
\mathcal{V}^+(x)- \mathcal{V}^-(x)=\left(\begin{array}{c}
0\\
0\\
0\\
\phi^A\delta^A P_A+\omega^A L_A+\phi^A_T\delta^A_T T_A
+a^B \left((1-\zeta) I_C+\zeta I_G\right)\\
\phi^C\delta^C P_C+\omega^C L_C+\phi^C_T\delta^C_T T_C+ a^A (1-\zeta) I_A\\
\phi^C\delta^C P_G+\omega^C L_G+\phi^G_T\delta^C_T T_G+ a^A\zeta I_A\\
(1-\phi^A)\delta^A P_A+(1-\phi^A_T)\delta^A_T T_A
+a^B \left((1-\zeta) L_C+\zeta L_G\right)\\
(1-\phi^C)\delta^C P_C+(1-\phi^C_T)\delta^C_T T_C+a^A (1-\zeta) L_A\\
(1-\phi^C)\delta^C P_G+(1-\phi^G_T)\delta^C_T T_G+a^A \zeta L_A\\
\tau^A k^A I_A+a^B \left((1-\zeta) T_C+\zeta T_G\right)\\
\tau^C k^C I_C+a^A (1-\zeta) T_A\\
\tau^C k^C I_G+a^A \zeta T_A\\
\eta^A +a^B \left((1-\zeta) S_C+\zeta S_G\right)\\
\eta^C +a^A (1-\zeta) S_A\\
\eta^C +a^A \zeta S_A\\
\end{array}\right)-\left(\begin{array}{c}
\left(\delta^A+M_A\right)P_A\\
\left(\delta^C+M_C\right)P_C\\
\left(\delta^C+M_G\right)P_G\\
\left(\tau^A k^A +M_A\right)I_A\\
\left(\tau^C k^C +M_C\right)I_C\\
\left(\tau^C k^C +M_G\right)I_G\\
\left(\sigma^A\lambda_A+\omega^A+M_A\right)L_A\\
\left(\sigma^C\lambda_C+\omega^C+M_C\right)L_C\\
\left(\sigma^C\lambda_G+\omega^C+M_G\right)L_G\\
\left(\delta^A_T+M_A\right)T_A\\
\left(\delta^C_T+M_C\right)T_C\\
\left(\delta^C_T+M_G\right)T_G\\
\left(\lambda_A +M_A\right)S_A\\
\left(\lambda_C +M_C\right)S_C\\
\left(\lambda_G +M_G\right)S_G\\
\end{array}\right).$$
}
Note that $\mathcal{F}_{1-3}$ denotes the entries of $\mathcal{F}$ from $1$
to $3$. Then $\mathcal{F}$ and $\mathcal{V}= \mathcal{V}^+- \mathcal{V}^-$
satisfy the following assumptions:
\begin{itemize}
\item[$(A_1)$] if $x\geq0$, then $\mathcal{F}(x)$, $\mathcal{V}^+(x)$,
$\mathcal{V}^-(x)\geq 0$ (each function represents a direct transfer of individuals);
\item[$(A_2)$] if $x_i=0$, then $\mathcal{V}_i^-(x)=0$
(if the compartment is empty, then there cannot be out-transfers of individuals);
\item[$(A_3)$] $\mathcal{F}_i(x)=0$ for $i>12$;
\item[$(A_4)$] if $x\in X_s$, then $\mathcal{F}_i(x)=0$ and $\mathcal{V}_i^+(x)=0$
for $1\leq i\leq 12$ (if the population is free of disease,
then it will remain free of disease);
\item[$(A_5)$] when $\mathcal{F}(x)=0$ we have that $DX(x_0)$ is a Hurwitz
matrix, i.e., all eigenvalues have negative real part
(the equilibrium point $x_0$ is asymptotically stable).
\end{itemize}
Only assumption $(A_5)$ creates some difficulty, since the other assumptions
are evident. We numerically checked~$(A_5)$ (in all calculations made)
using the Routh--Hurwitz criterion, which states that the matrix $A=D X(x_0)$
is Hurwitz if and only if all the principal subdeterminants, of a special matrix
constructed with the coefficients of the characteristic polynomial of $A$,
are all strictly positive.

By Lemma~1 in~\cite{Driessche-Watmough}, the derivatives $D\mathcal{F}(x_0)$
and $D\mathcal{V}(x_0)$ are partitioned as
$$
D\mathcal{F}(x_0)=\left(\begin{array}{cc}F & 0\\0 & 0\end{array}\right)
\quad\mbox{ and }\quad
D\mathcal{V}(x_0)=\left(\begin{array}{cc}V & 0\\J_3 & J_4\end{array}\right),
$$
where $F$ and $V$ are $m\times m$-matrices. Hence,
we have $F_{i,j}(x)=0$, if $i>m$ or $j>m$, and
$$
F_{1-6,1-6}=\left(\begin{array}{cccccc}
0 & a^B (1-\zeta) & a^B \zeta & \beta^A\nu^A & 0 & 0\\
a^A(1-\zeta) & 0 & 0 & 0 & \beta^C\nu^C & 0\\
a^A\zeta & 0 & 0 & 0 & 0 & \beta^G\nu^G\\
0 & 0 & 0 & 0 & 0 & 0\\
0 & 0 & 0 & 0 & 0 & 0\\
\end{array}\right).
$$
The critical threshold function $R_0^T$ is then given as the spectral radius
of the matrix $A=FV^{-1}$. We have that $A$ has all entries zero except
\myeqAN{
A_{1,i}&=  a^B (1-\zeta)V_{2,i}^{-1}+a^B \zeta V_{3,i}^{-1}
+\beta^A\nu^A V_{4,i}^{-1},\\
A_{2,i}&= a^A (1-\zeta) V_{1,i}^{-1}+\beta^C\nu^C V_{5,i}^{-1},\\
A_{3,i}&= a^A \zeta V_{1,i}^{-1}+\beta^G\nu^G V_{6,i}^{-1}.
}
Considering the algebraic complexity of computing the spectral radius of $A$,
in the next subsection we proceed numerically by understanding $R_0$
from the variation of the parameters.

% ------------------------------------------

\subsection{Sensitivity analysis: numerical simulations}

The values of the parameters $\beta$, $\nu$, $\mu$, $\delta$, $\phi$, $\sigma$,
$\omega$, $\tau$, $k$, $\delta_T$ and $\phi_T$ estimated for Portugal, are based
on the values proposed in \cite{TBportugalGomesRodrigues}, as well as the initial
conditions $N(0)$, $\TS(0)$, $\TP(0)$, $\TL(0)$, $\TI(0)$, $\TT(0)$. We assume
that the Portuguese total population will decrease ($\eta < \mu N$), based on
the projections for resident population in Portugal from \emph{Statistics Portugal}
\cite{INE2014} and the value for TB induced death that comes from \cite{Styblo_1991}.

We assume that the reference value for the transmission coefficient in Angola
is $\beta = 150$ based on \cite{noticiasAngola2}. According to the World Bank,
the natural death rate in Angola is equal to $\mu = 1/51 \,  yrs^{-1}$
\cite{worldbankdT}. The value for the TB induced death rate is based
on \cite{Styblo_1991}. The proportion of pulmonary TB cases in Angola is equal
to $\nu = 0.937$ and the fraction of treatment default and failure for individuals
under treatment is equal to $\phi_T = 0.219$ \cite{noticiasAngola1}. We assume
that the reinfection factor $\sigma$ in Angola takes the value proposed
in \cite{TBportugalGomesRodrigues}. According to WHO, the proportion
of detected cases in a year is equal to $k=0.79$ \cite{WHO:TB:report:2013}.
The rate at which infectious individuals enter treatment is estimated to be
$\tau = 2.13 \, yrs^{-1}$. The values of the parameters $\delta$, $\phi$,
$\omega$ and $\delta_T$ are taken from \cite{TBportugalGomesRodrigues}.
The recruitment rate value $\eta = 1287900$ is based on the population
projections from Population Reference Bureau \cite{PopulationRankings}.
The initial conditions $N(0)$, $\TS(0)$, $\TP(0)$, $\TL(0)$, $\TI(0)$,
$\TT(0)$ are based on data from \cite{SilvaTorresTBAngola,wikiAngola,noticiasAngola2}.
All previous values are resumed in Table~\ref{table:parameters:PT:AN}.
%-----------------------------
\begin{table}[!htb]
\centering
\begin{tabular}{|l | l | l | l |}
\hline
{\scriptsize{Symbol}} & {\scriptsize{Description}}  & {\scriptsize{Portugal}}
&  {\scriptsize{Angola}}\\
\hline
{\scriptsize{$\beta$}} & {\scriptsize{Transmission coefficient}}
& {\scriptsize{variable ($72.358 \, yrs^{-1}$) }}
& {\scriptsize{variable ($150 \, yrs^{-1}$)}}\\
{\scriptsize{$\nu$}} & {\scriptsize{Proportion of pulmonary TB cases}}
& {\scriptsize{$0.75$}} & {\scriptsize{$0.937$}}\\
{\scriptsize{$\mu$}} & {\scriptsize{Natural death rate}}
& {\scriptsize{$1/80 \, yrs^{-1}$}} & {\scriptsize{$1/51 \, yrs^{-1}$}}\\
{\scriptsize{$\delta$}} & {\scriptsize{Rate at which individuals leave P compartment}}
& {\scriptsize{$2 \, yrs^{-1}$}} & {\scriptsize{$2 \, yrs^{-1}$}}\\
{\scriptsize{$\phi$}} & {\scriptsize{Fraction of infected population developing active TB}}
& {\scriptsize{$0.05$}} & {\scriptsize{$0.05$}}\\
{\scriptsize{$\sigma$}} & {\scriptsize{Reinfection (exogenous) factor for latent}}
& {\scriptsize{$0.5$}} & {\scriptsize{$0.5$}}\\
{\scriptsize{$\omega$}} & {\scriptsize{Rate of endogenous reactivation for latent infections}}
& {\scriptsize{$0.0003 \, yrs^{-1}$}} & {\scriptsize{$0.0003 \, yrs^{-1}$}}\\
{\scriptsize{$\tau$}} & {\scriptsize{Rate at which infectious individuals enter treatment}}
& {\scriptsize{$4.26 \, yrs^{-1}$}} & {\scriptsize{$2.13 \, yrs^{-1}$}}\\
{\scriptsize{$k$}} & {\scriptsize{Proportion of detected cases in a year}}
& {\scriptsize{$0.87$}} & {\scriptsize{$0.79$}}\\
{\scriptsize{$\delta_T$}} & {\scriptsize{Inverse of treatment length}}
& {\scriptsize{$1.36 \, yrs^{-1}$ }} & {\scriptsize{$1.36 \, yrs^{-1}$}}\\
{\scriptsize{$\phi_T$}} & {\scriptsize{Fraction of treatment default and failure}}
& {\scriptsize{$0.04$}} & {\scriptsize{$0.219$}}\\
{\scriptsize{$\eta$}} & {\scriptsize{Recruitment rate for Portugal}}
& {\scriptsize{$78672$ }} & {\scriptsize{$1287900$ }} \\
{\scriptsize{$d_T$}} & {\scriptsize{TB induced death rate for Portugal}}
& {\scriptsize{$1/5 \, yrs^{-1}$ }} & {\scriptsize{$1/8 \, yrs^{-1}$}} \\
{\scriptsize{$N(0)$}} & {\scriptsize{Initial total population}}
& {\scriptsize{10560000}} & {\scriptsize{24300000}}\\
{\scriptsize{$\TS(0)$}} & {\scriptsize{Initial susceptible population}}
& {\scriptsize{$8947300$}} & {\scriptsize{$9618729$}}\\
{\scriptsize{$\TP(0)$}} & {\scriptsize{Initial primary infected with TB population}}
& {\scriptsize{$11000$}} & {\scriptsize{$24300$}}\\
{\scriptsize{$\TI(0)$}} & {\scriptsize{Initial actively infected (and infectious) population}}
& {\scriptsize{$500$}} & {\scriptsize{$16164$}}\\
{\scriptsize{$\TL(0)$}} & {\scriptsize{Initial latent infected population}}
& {\scriptsize{$1600000$}} & {\scriptsize{$14580000$}}\\
{\scriptsize{$\TT(0)$}} & {\scriptsize{Initial under treatment population}}
& {\scriptsize{$1200$}} & {\scriptsize{$60807$}}\\
\hline
\end{tabular}
\caption{Estimated parameters and initial conditions values for Portugal and Angola.}
\label{table:parameters:PT:AN}
\end{table}
%-------------------------------------------

If we firstly keep all parameters fixed (see Table~\ref{table:parameters:PT:AN}),
we have
$$
R_0^T= \numD{6.359799999}.
$$
Then we vary one of the parameters $\beta^A$, $\beta^C$, $\beta^G$, $k^C$,
$\phi^G_T$, $a^A$, $a^B$, or $\zeta$ in the ranges
$$
\begin{array}{ccc}
150(1-\theta)\leq\beta^A\leq 150(1+\theta), &
72.358(1-\theta)\leq\beta^C\leq 72.358(1+\theta), \\
\beta^C=72.358\leq \beta^G\leq 150=\beta^A, &
0.87(1-\theta)\leq k^C\leq 0.87(1+\theta), \\
\phi_T^C= 0.04, \leq \phi_T^G\leq 0.219=\phi_T^A, &
0\leq a^A\leq 0.1, \\
0\leq a^B \leq 0.1, &
0\leq \zeta\leq 1, \\
\end{array}
$$
where $\theta=0.2$. Each simulation gives a curve $x\mapsto R_0^T(x)$,
where $x$ is one of the above parameters, for which we find a best fitting
curve in one of the models
\myeq{fitmodels}{
P_n(x)=a_0 +a_1 x+ a_2 x^2+\dots+a_n x^n, \:\:n\in\{0,\dots,99\},
\quad\mbox{ and }\quad \frac{a_0+a_1 x + a_2 x^2}{b_0+b_1 x + b_2 x^2},
}
for some constants $a_0,\dots, a_n,b_0, b_1, b_2\in\bkR^N$.
%--------------------------------
\begin{table}[!htb]
\centering
\begin{tabular}{| c | c | c | c |}
\hline
{\scriptsize Parameter} & {\scriptsize Type} & {\scriptsize Curve Fitting}
& {\scriptsize $\log_{10}(SQR)$}\\
\hline
{\scriptsize $\beta^A$} & {\scriptsize best} & {\scriptsize $R_0^T
=\numD{0.009021048}+\numD{0.0422905212}\, \beta^A
+\numD{4.39929508e-7}\, (\beta^A)^2$} & \ \\
\ & \ & {\scriptsize $\numD{-7.99262125e-10}\, (\beta^A)^3$}
& {\scriptsize $\numD{-5.278627639}$}\\
\ & {\scriptsize as in $R_0^A$} & {\scriptsize $R_0^T=\numD{.042398419}\, \beta^A$}
&  {\scriptsize $\numD{-2.2700918}$}\\
\ & \ & \ & \ \\
{\scriptsize $\beta^C$} & {\scriptsize best} & {\scriptsize $R_0^T
=\numD{6.3577692}+\numD{2.2968396e-005}\, \beta^C
+\numD{6.545004e-008}\, (\beta^C)^2$} & \ \\
\ & \ & {\scriptsize $+\numD{6.933227073e-12}\, (\beta^C)^3
+\numD{8.83463551e-13}\, (\beta^C)^4$} & {\scriptsize \numD{-7.055330529}} \\
\ & {\scriptsize as in $R_0^C$} & {\scriptsize $R_0^T
=\numD{.08671477}\, \beta^C$} &  {\scriptsize $\numD{.8692731}$}\\
\ & \ & \ & \ \\
{\scriptsize $\beta^G$} & {\scriptsize best}& {\footnotesize $R_0^T
=\frac{\numD{-3639.13063363}+\numD{1285.78172120}\, \beta^G
-\numD{4.168861981}\, (\beta^G)^2}{\numD{-572.32092506}+\numD{202.21340477}\,
\beta^G -\numD{0.65585384}\, (\beta^G)^2}$} & {\scriptsize $\numD{-7.0129799}$} \\
\ & {\scriptsize as in $R_0^G$} & {\scriptsize $R_0^T=\numD{.0551009}\, \beta^G$}
&  {\scriptsize $\numD{1.097186}$}\\
\ & \ & \ & \ \\
{\scriptsize $k^C$} & {\scriptsize best} & {\scriptsize $R_0^T=P_{89}(k^C)$}
& {\scriptsize $\numD{-7.058232835}$}\\
\ & {\scriptsize not best} & {\footnotesize $R_0^T=\frac{\numD{-27.075657}
+\numD{53.2692494}\,k^C+\numD{215.9660609}\, (k^C)^2}{\numD{-4.2640340}
+\numD{8.36667709}\, k^C +\numD{33.97750881}\, (k^C)^2}$}
& {\scriptsize $\numD{-6.6542336}$}\\
\ & \ & \ & \ \\
{\scriptsize $\phi_T^G$}  & {\scriptsize best} & {\scriptsize $R_0^T=P_{21}(\phi_T^G)$}
& {\scriptsize $\numD{-7.007144237}$}\\
\  & {\scriptsize not best} & {\footnotesize $R_0^T=\frac{\numD{1103.5908788}
-\numD{1522.87667500}\, \phi_T^G}{\numD{173.537338}-\numD{239.505221}\, \phi_T^G}$}
& {\scriptsize $\numD{-7.002127}$}\\
\ & \ & \ & \ \\
{\scriptsize $a^A$}  & {\scriptsize best} & {\footnotesize $R_0^T
=\frac{\numD{38.07598473}+\numD{2747.0900415}\, a^A+\numD{42528.207079}\,
(a^A)^2}{\numD{6.0248383}+\numD{419.985378}\, a^A+\numD{6317.049255}\, (a^A)^2}$}
& {\scriptsize $\numD{-4.27964}$}\\
\ & \ & \ & \ \\
{\scriptsize $a^B$} & {\scriptsize best} & {\scriptsize $R_0^T
=\numD{6.78217128}-\numD{77.5566782}\,a^B +\numD{2571.6270531}\, (a^B)^2$}  & \ \\
\ & \ & {\scriptsize $-\numD{68307.97202742} \, (a^B)^3+ \numD{1194841.268572}\,
(a^B)^4- \numD{12711602.9588141}\, (a^B)^5$} & \ \\
\ & \ & {\scriptsize $+\numD{73922994.9730162}\,(a^B)^6-\numD{179395541.509671}\,(a^B)^7$}
& {\scriptsize $\numD{-2.1925449}$}\\
\ & \ & \ & \ \\
{\scriptsize $\zeta$} & {\scriptsize best} & {\scriptsize $R_0^T=P_{91}(\zeta)$}
& {\scriptsize $\numD{-7.11446753}$}\\
\ & {\scriptsize not best} & {\scriptsize $R_0^T=\numD{6.383321621}
-\numD{.1147570445}\, \zeta+\numD{.1218747529}\, \zeta^2$}
& {\scriptsize $\numD{-3.0198063}$}\\
\hline
\end{tabular}
\caption{Curve fitting of $R_0^T$.}
\label{R0TFit}
\end{table}

Table~\ref{R0TFit} shows several curve fittings for the map $x\mapsto R_0^T(x)$.
By ``best fitting'' we mean a model, chosen between the above
models~\myref{fitmodels}, where the square root of the sum of squares of the
residuals $SQR=\sqrt{\sum_i r_i^2}$ has a minimum value or is smaller than the
number of significant digits in determining $R_0^T$, i.e., $10^{-8}$. The same
procedure applied to $R_0^A, R_0^C, R_0^G$ gave results compatible
with the analytic formula~\myref{eqR0s}.

% ------------------------------------------

\subsubsection{Variation of the TB transmission rates
(i.e., changing $\beta^A$, $\beta^C$ and $\beta^G$)}

\begin{figure}[!htb]
\begin{subfigure}[b]{0.33\textwidth}
\centering
\includegraphics[width=\textwidth]{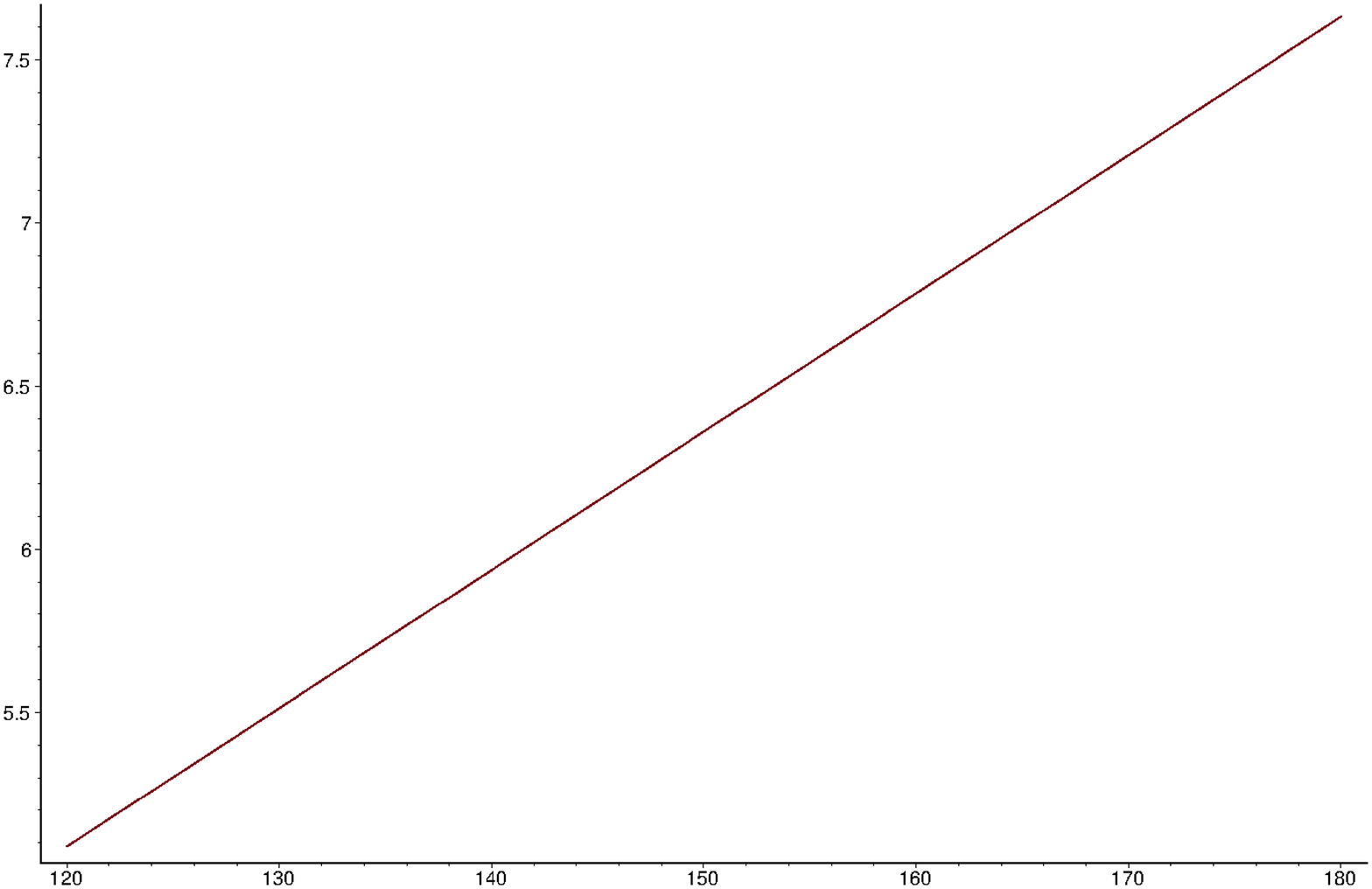}
\SFcaption{$R_0^T \in [5.1, 7.6]$ vs $\beta^A \in [120, 180]$}
\end{subfigure}
\begin{subfigure}[b]{0.33\textwidth}
\centering
\includegraphics[width=\textwidth]{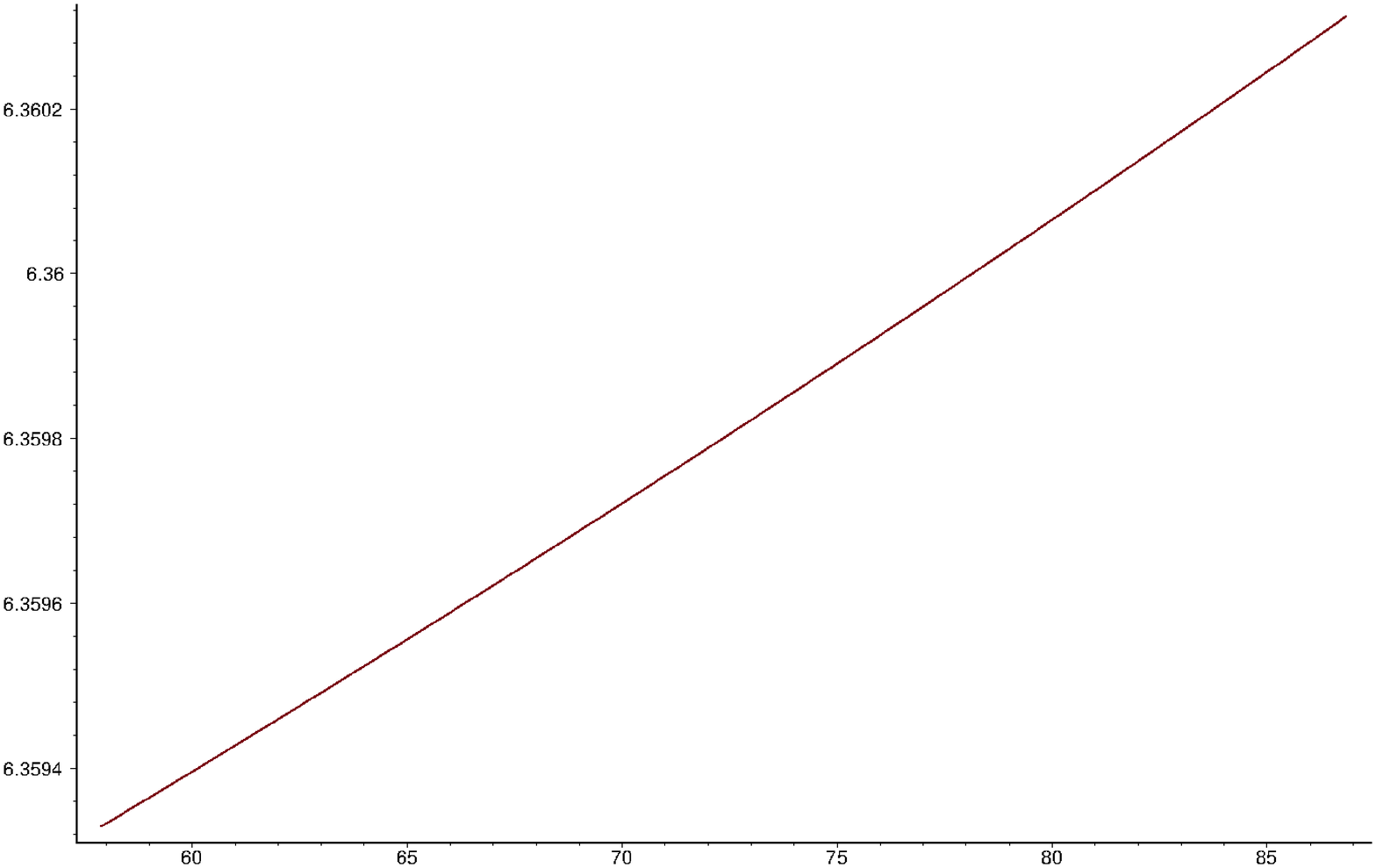}
\SFcaption{$R_0^T \in [6.3593, 6.3603]$ vs $\beta^C \in [78, 87]$}
\end{subfigure}
\begin{subfigure}[b]{0.33\textwidth}
\centering
\includegraphics[width=\textwidth]{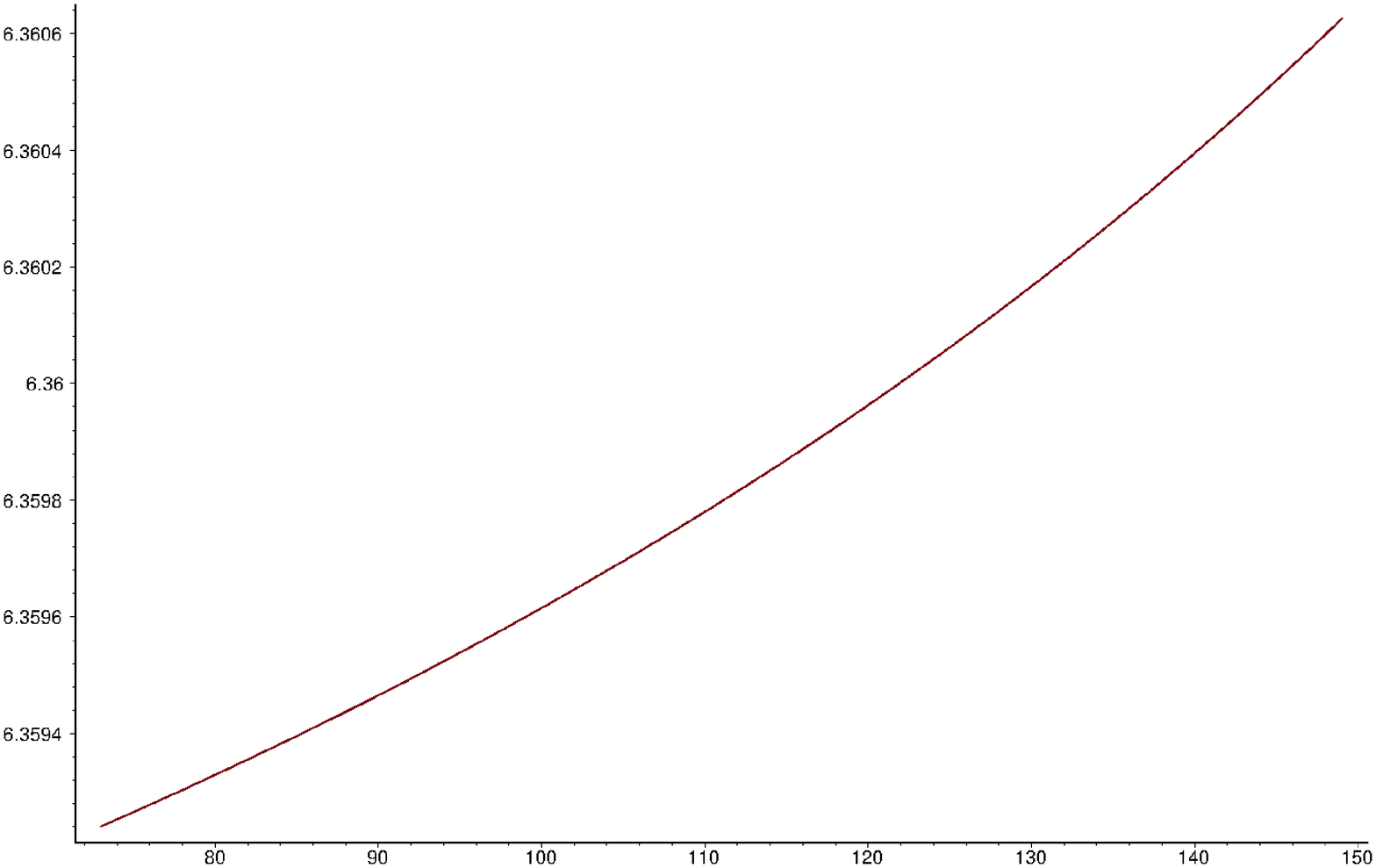}
\SFcaption{$R_0^T \in [6.3592, 6.3607]$ vs $\beta^G \in [70,150]$}
\end{subfigure}
\caption{$R_0^T$ when varying $\beta^A$, $\beta^C$ and $\beta^G$, respectively.}
\label{R0Tbeta}
\end{figure}

A variation of $20\%$ in the value of $\beta^A$ implies a variation of
approximately $20\%$ to $R_0^T$. However, the same variation of $20\%$ in the
values of $\beta^C$ and $\beta^G$ affects $R_0^T$ less than $1\%$. Contrary
to~\myref{eqR0s}, the parameters $\beta^A, \beta^C, \beta^G$ do not appear
linearly in the calculation of $R_0^T$, although locally look similar
to an affine function, see Fig.~\ref{R0Tbeta}.

\begin{figure}[!htb]
\begin{subfigure}[b]{0.5\textwidth}
\centering
\includegraphics[width=\textwidth]{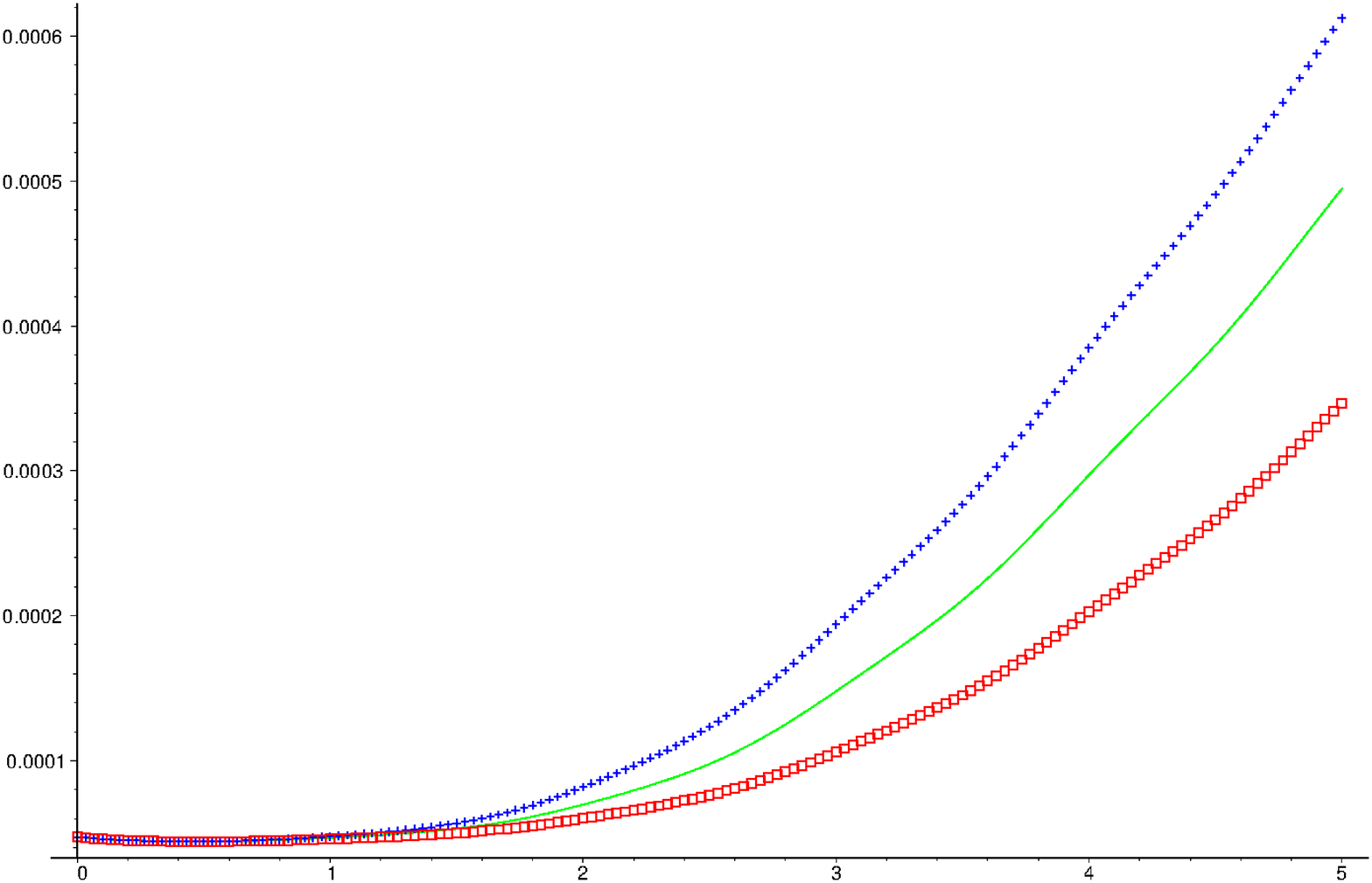}
\SFcaption{$I_C(t) \in [0, 0.0006]$ vs $t \in [0, 5]$}
\end{subfigure}
\hspace*{0.1cm}
\begin{subfigure}[b]{0.5\textwidth}
\centering
\includegraphics[width=\textwidth]{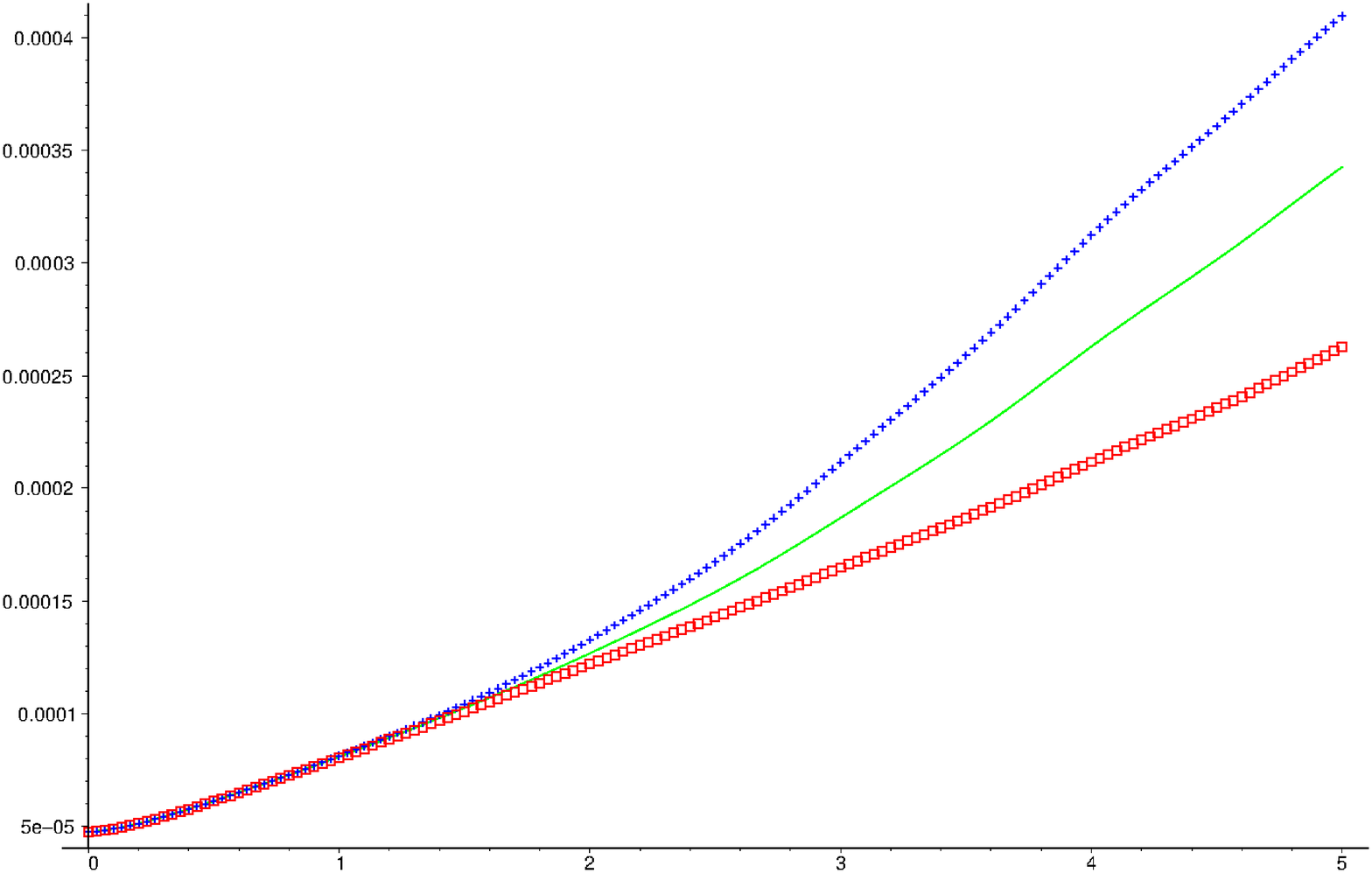}
\SFcaption{$I_G(t) \in [0, 0.0004]$ vs $t \in [0, 5]$}
\end{subfigure}
\caption{$I_C(t)$ and $I_G(t)$ when varying $\beta^A$
(box: $\beta^A=120$, solid: $\beta^A=150$, cross: $\beta^A=180$).}
\label{CIGI_ABETA}
\end{figure}

The variation of $\beta^A$ has also a significative impact on the community and
the host country, namely, in the number of infected and infectious individuals
after $5$ years, see Fig.~\ref{CIGI_ABETA}. Defining
$I_X(t, s)=\left.I_C(t)\right|_{\beta^A=s}$ with $X\in\{C,G\}$, we have
$$
\frac{I_C(5,180)}{I_C(5,150)}\approx 1.24,
\quad \frac{I_C(5,120)}{I_C(5,150)}\approx 0.70,
\quad\frac{I_G(5,180)}{I_G(5,150)}\approx 1.20,
\quad \frac{I_G(5,120)}{I_G(5,150)}\approx 0.77.
$$
An increase (decrease) of $20\%$ in $\beta^A$ implies a 5 years increase of
approximately $20\%$ (decrease of $30\%$) in $I_C$ and $I_G$, respectively.
This enforce the importance of additional effort to treat TB in countries
with high TB incidence, not only because of their population health improvement,
but also because of the implications on the health of individuals
in other host countries.

% ------------------------------------------

\subsubsection{Variation in the transfer of individuals (i.e., changing $a^A$ and $a^B$)}

\begin{figure}[!htb]
\begin{subfigure}[b]{0.5\textwidth}
\centering
\includegraphics[width=\textwidth]{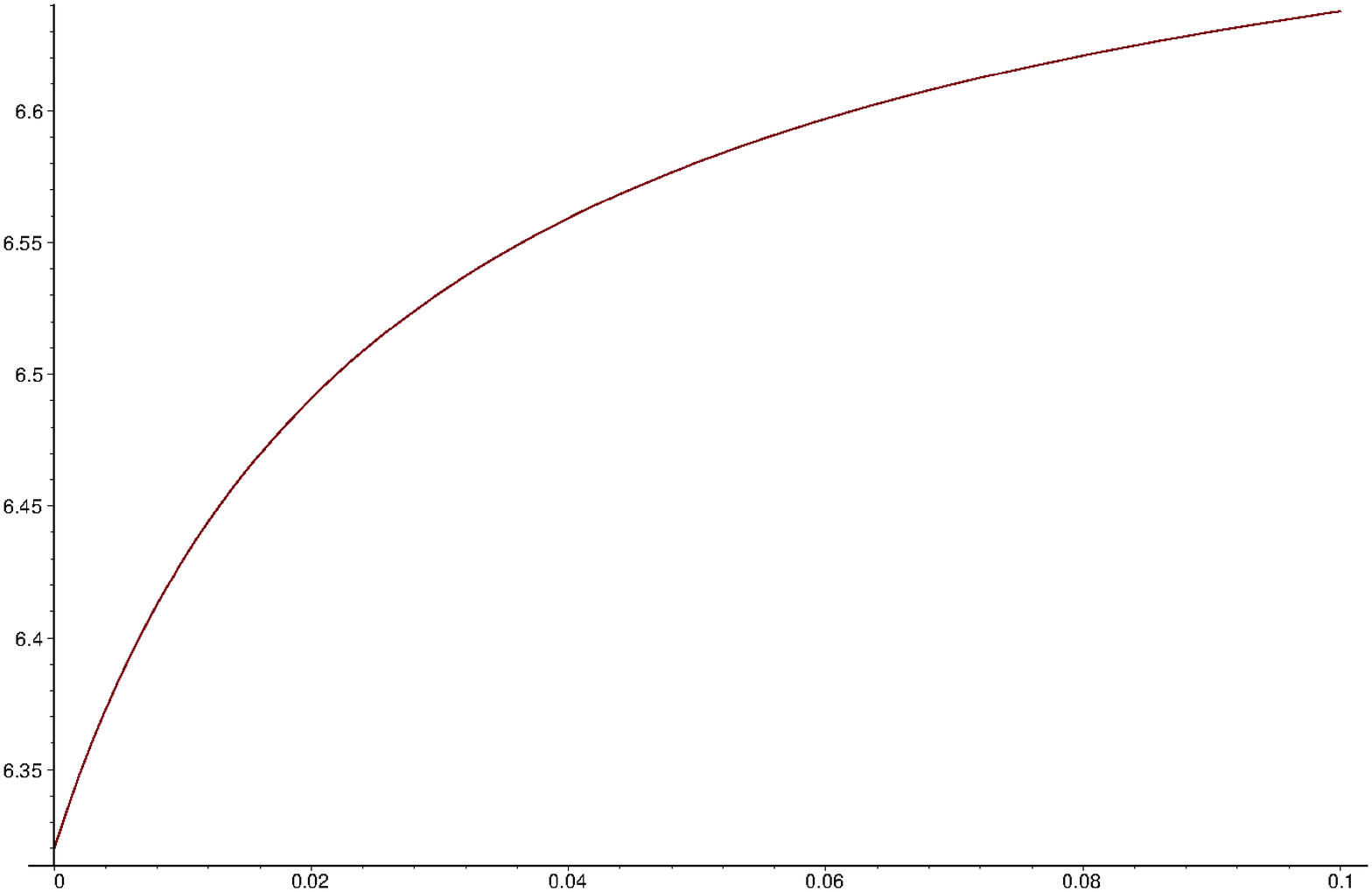}
\SFcaption{$R_0^T \in [6.3, 6.7]$ vs $a^A \in [0, 0.1]$}
\end{subfigure}
\begin{subfigure}[b]{0.5\textwidth}
\centering
\includegraphics[width=\textwidth]{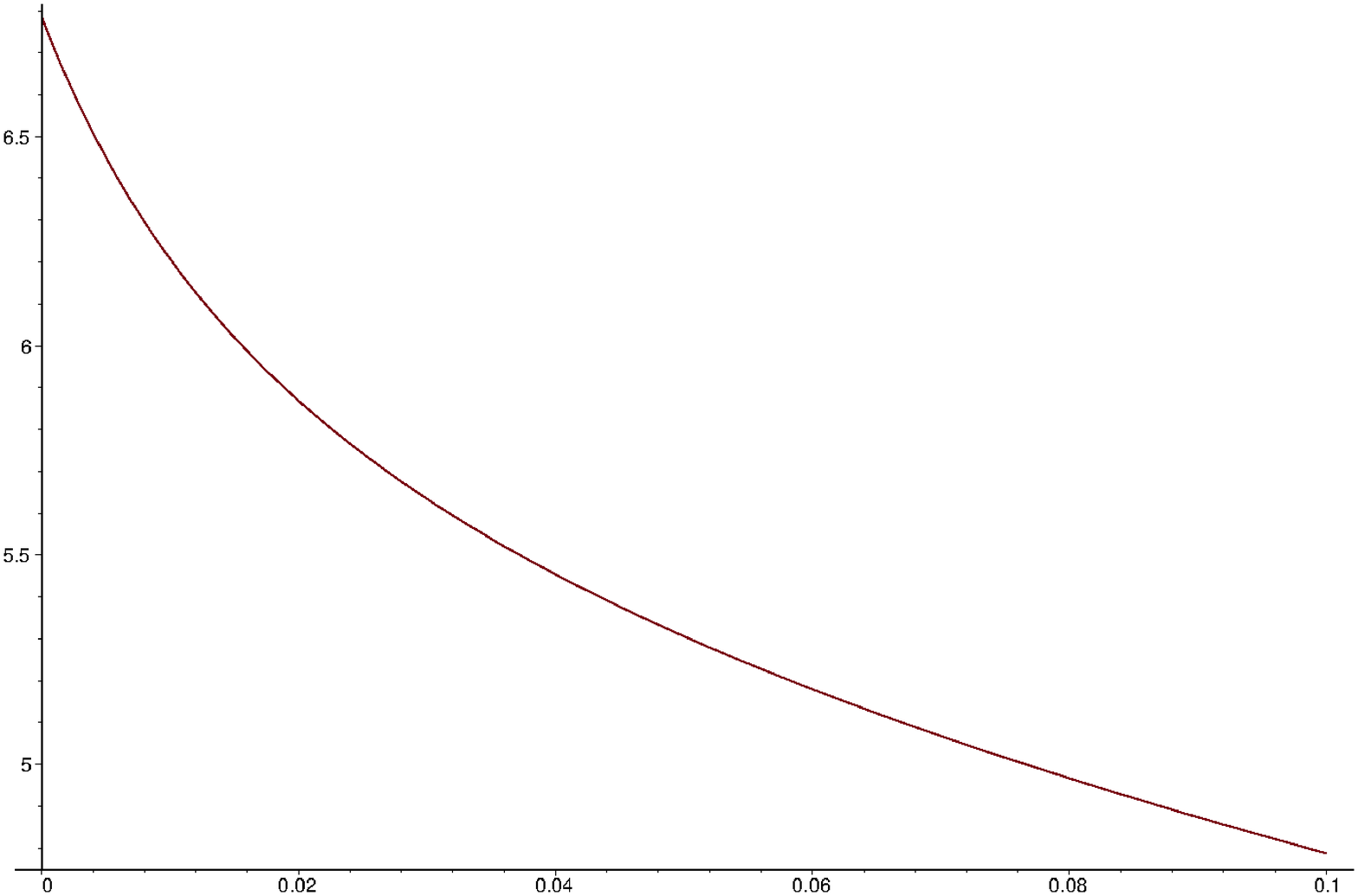}
\SFcaption{$R_0^T \in [4.8, 6.8]$ vs $a^B \in [0, 0.1]$}
\end{subfigure}
\caption{$R_0^T$ when varying $a^A$ and $a^B$, respectively.}
\label{R0Ta}
\end{figure}

The transfer of individuals between (A) and (C)+(G) (i.e., (B)) is determined by the
functions $\gamma_A(t)$ and $\gamma_B(t)$, which are here assumed to be equal
to the parameters $a^A$ and $a^B$. From Fig.~\ref{R0Ta} it is clear, as expected,
that an increment on the flux of individuals moving from areas of lower
TB incidence to areas of higher TB incidence reduces $R_0^T$ and,
on the contrary, an increment in the flux of individuals moving from areas
of high TB incidence to areas of lower TB incidence increases $R_0^T$.
Note that $R_0^T$ grows very fast for smaller values of $a^A$ and then tends
to stabilize with the flux of persons coming from the high incidence TB area.

\begin{figure}[!htb]
\begin{subfigure}[b]{0.5\textwidth}
\centering
\includegraphics[width=\textwidth]{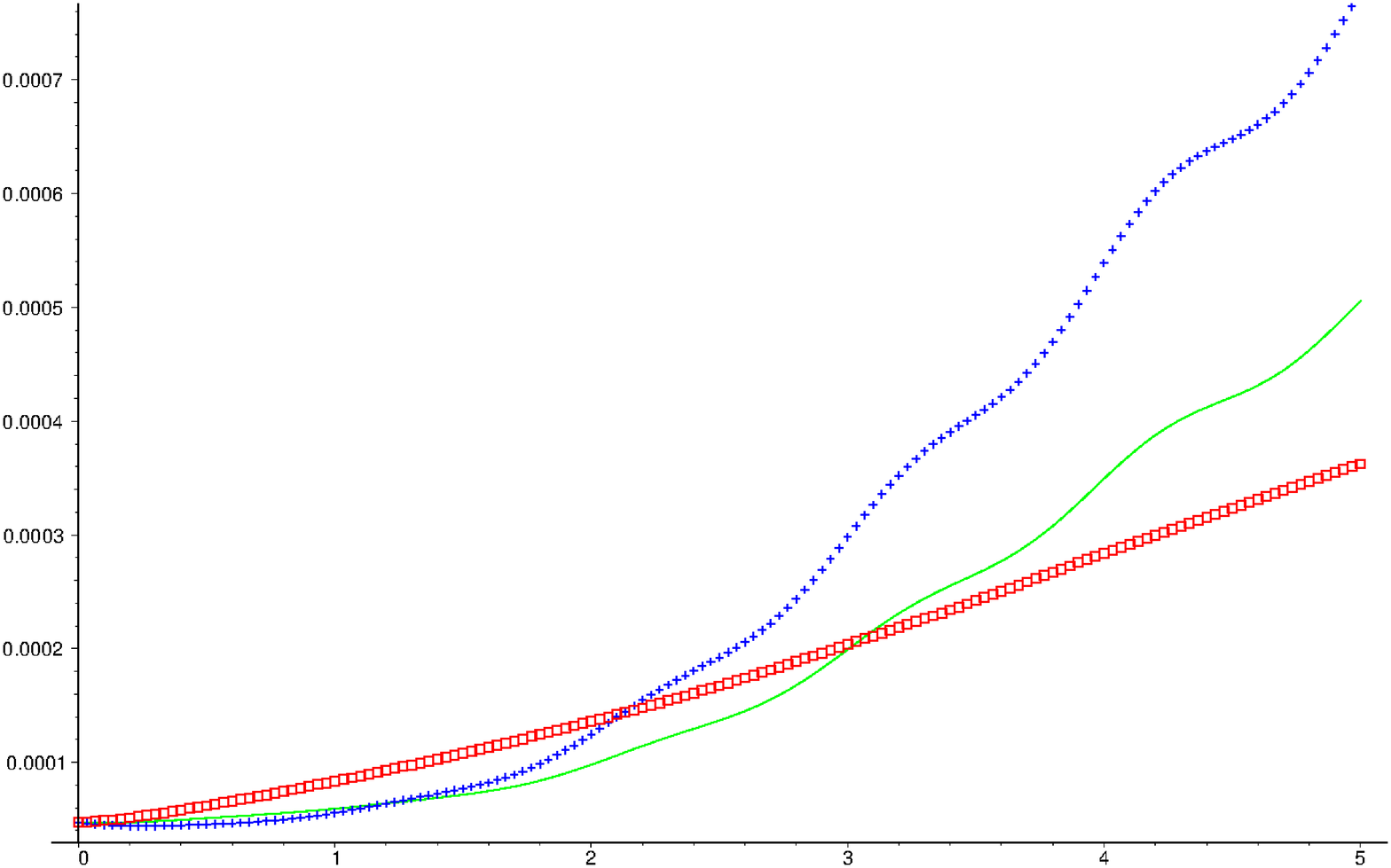}
\SFcaption{$I_G(t) \in [0, 0.0008]$ vs $t \in [0, 5]$}
\end{subfigure}
\begin{subfigure}[b]{0.5\textwidth}
\centering
\includegraphics[width=\textwidth]{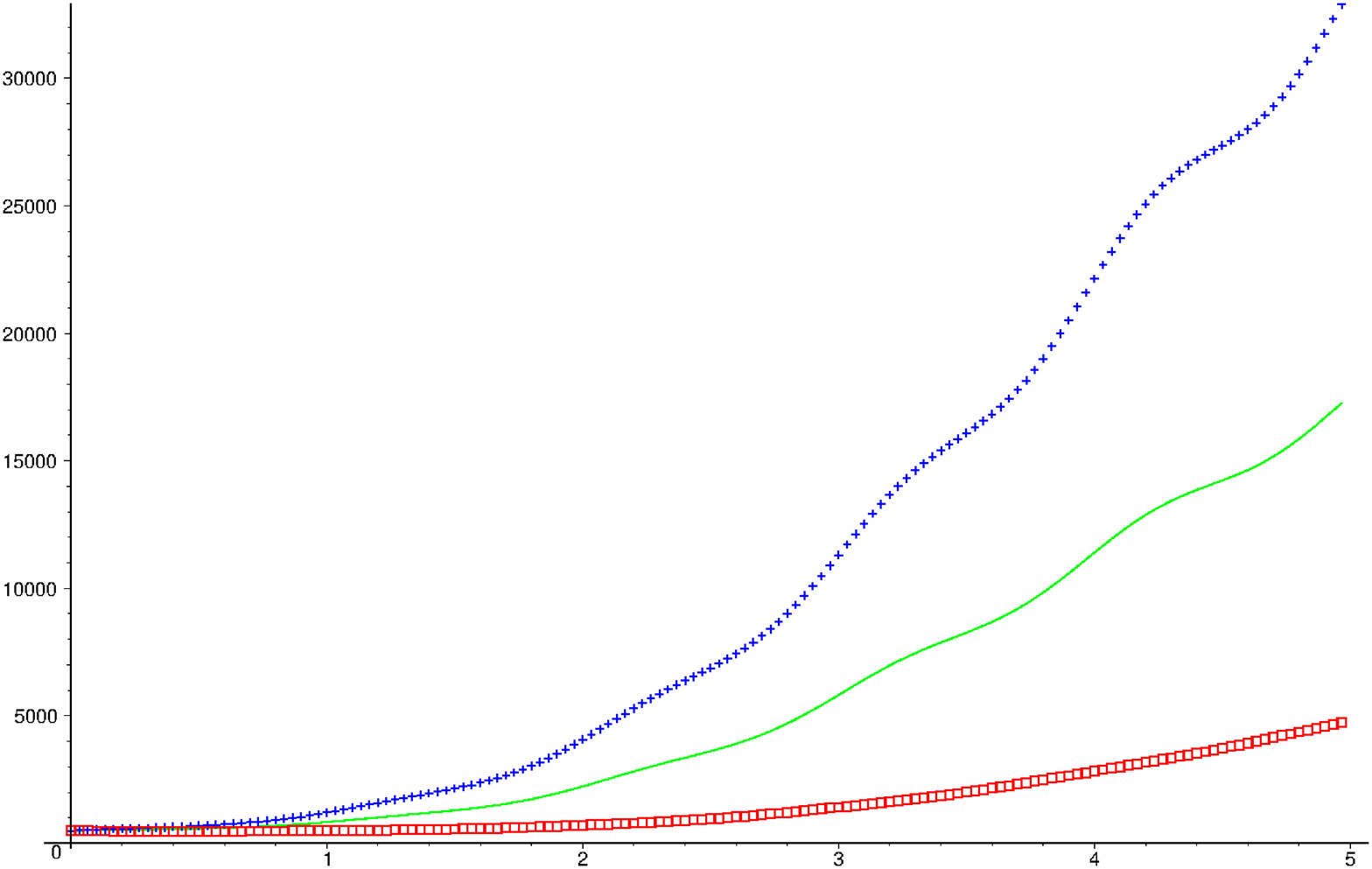}
\SFcaption{$\mathcal{N} \in [0, 35000]$ vs $t \in [0, 5]$}
\end{subfigure}
\caption{$I_G(t)$ and total number $\mathcal{N}$ of infected individuals
in $(C)+(G)$ when varying $a^A$ (box: $a^A=0$, solid: $a^A=0.05$, cross: $a^A=0.1$).}
\label{R0Tb}
\end{figure}

An interesting phenomena when varying $a^A$ appears in the variable $I_G$,
i.e., the number of infected individuals in (G) (the community), see Fig.~\ref{R0Tb}.
It tells us that it is better for the community to have some moderate exchange
of persons with the high incidence TB region. Such behavior and its reverse,
after some time, seems to be related to the chosen value of $\zeta$
(discussed in the next subsection). It also imply that a careful study of the
seasonality distribution of persons traveling between (A) and (B) may be more relevant
for (G) than expected a priori. On the host country viewpoint, such phenomena
is not noticed as one can see from the evolution of the total number of infected
individuals in the host country, i.e., $I_C(t)\,N_C(t)+I_G\,N_G(t)$, see Fig.~\ref{R0Tb}.

% ------------------------------------------

\subsubsection{About the ratio of individuals that stay in the community
versus spread in the host country (i.e., changing $\zeta$)}

\begin{figure}[ht]
\begin{subfigure}[b]{0.5\textwidth}
\centering
\includegraphics[width=\textwidth]{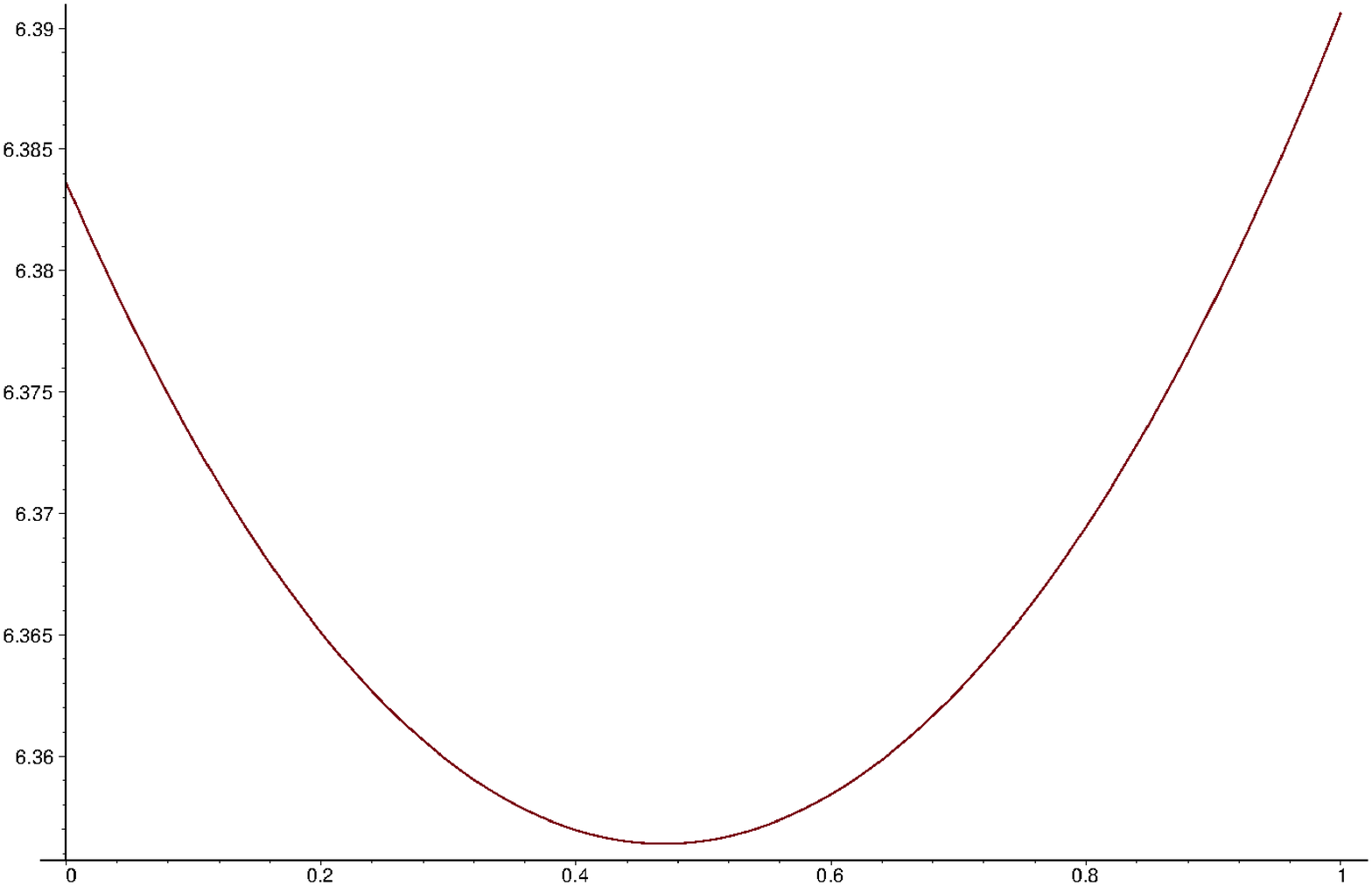}
\SFcaption{$R_0^T \in [6.35, 6.39]$ vs $\zeta \in [0, 1]$}
\end{subfigure}
\hspace*{0.1cm}
\begin{subfigure}[b]{0.5\textwidth}
\centering
\includegraphics[width=\textwidth]{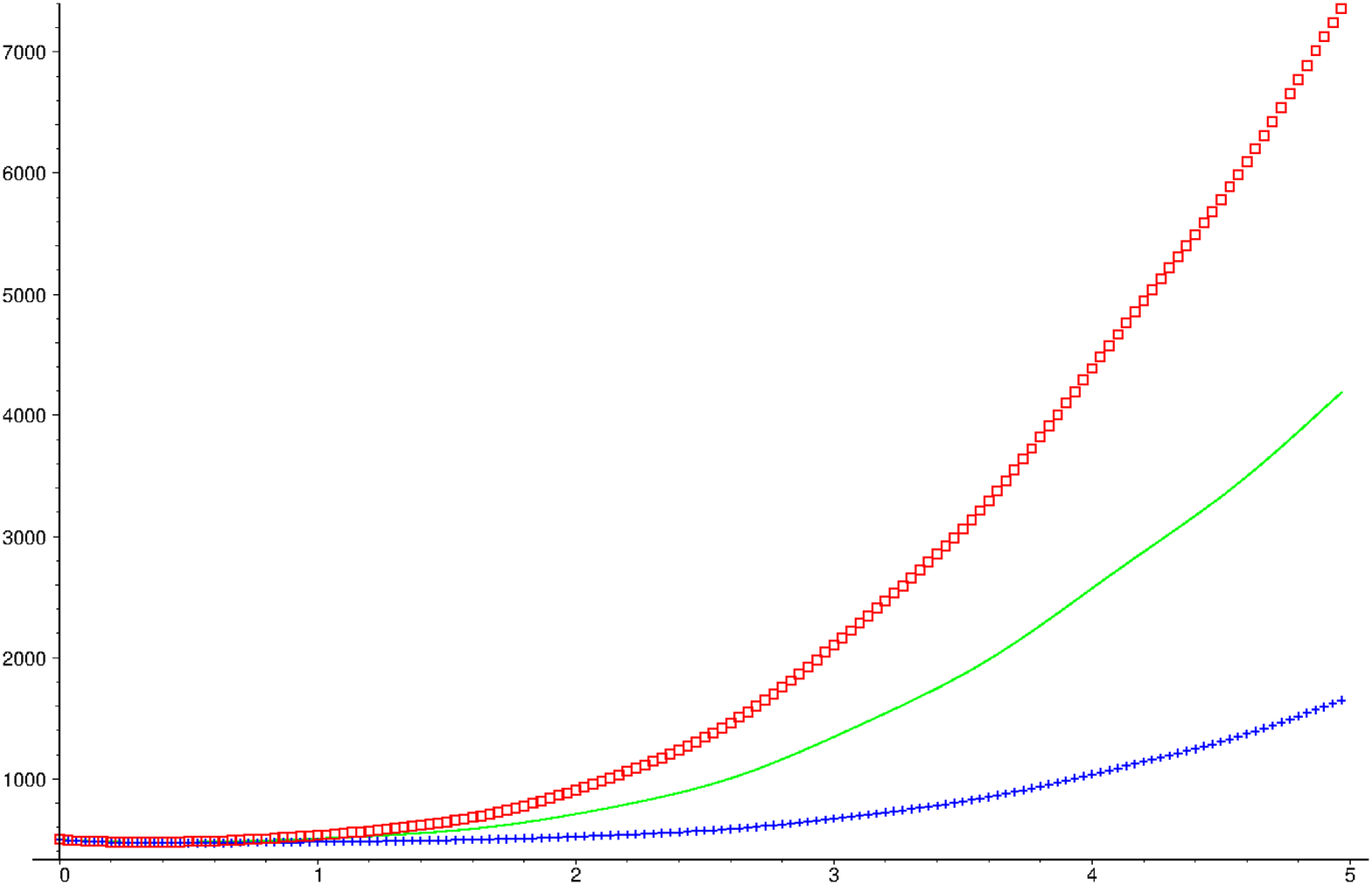}
\SFcaption{$\mathcal{H}(t) \in [0, 7500]$ vs $t \in [0, 5]$}
\end{subfigure}
\caption{$R_0^T$ versus $\zeta$ and total number 
$\mathcal{H}(t) = I_C(t)\,N_C(t)+I_G(t)\,N_G(t)$ of infected
individuals in the host country versus $t$ when changing $\zeta$ 
(box: $\zeta=0$; solid: $\zeta=0.5$; cross: $\zeta=1$).}
\label{FigZETA}
\end{figure}

In what follows we analyze the impact of the existence of a community
of immigrants coming from a high incidence TB area on the host country,
the country of origin and in the global situation.
Recall that $\zeta$ is the percentage of persons traveling that come/go
specifically to (G) versus the complementary (C). Hence, the situation $\zeta=0$
means that all persons traveling between Angola and Portugal all come/go
to (C) and none to (G). On the contrary, $\zeta=1$ means that all persons traveling
between Angola and Portugal all come/go to (G). From the analysis of
Fig.~\ref{FigZETA} (right), it is clear that the existence of a community
of immigrants coming from a high incidence TB area is convenient for the host
country in order to better control TB spread. Regarding the point of view
of Angola, a change in $\zeta$ is not significative as one can see,
in Table~\ref{Dvalues}, that $I_A$ is not affected by a change in $\zeta$.

On a global viewpoint, a change in $\zeta$ has a big impact on the reproduction
number $R_0^T$, see Fig.~\ref{FigZETA} (left), for which the existence
of communities turn to be also convenient. In fact, the function attains
a minimum value that can be estimated from the approximated fitting
by a parabolic function as
$$
R_0 (T ) = \numD{6.383321621} + \numD{0.114757045} x + \numD{0.121874753} x^2,
$$
see Table~\ref{R0TFit}. Hence, we may say that the optimal value for $\zeta$
is approximately
$$
\min_{0\leq\zeta\leq 1}R_0^T(\zeta)
= \frac{\numD{0.114757045}}{2\times \numD{0.121874753}}\approx 0.46.
$$

% ------------------------------------------

\section{Numerical results and discussion}
\label{sec:numeric}

Regarding the sensitivity analysis, we numerically simulated the
system~\myref{eqPA}--\myref{eqPGamma} by considering all parameters fixed except one chosen
parameter for which we consider three possible values according with
$$
\begin{array}{ll}
\beta^A \in \left\{150(1-\theta), 150, 150(1+\theta)\right\}, &
\beta^C \in \left\{72.358(1-\theta), 72.358, 72.358(1+\theta)\right\}, \\
\beta^G \in \left\{\beta^C, \frac{\beta^C+\beta^A}{2},\beta^A\right\}, &
k^C \in \left\{0.87(1-\theta), 0.87, 0.87(1+\theta)\right\}, \\
\phi_T^G \in \left\{\phi_T^C, \frac{\phi_T^C+\phi_T^A}{2}, \phi_T^A\right\}, &
a^A \in \left\{0, 0.05, 0.1\right\}, \\
a^B \in \left\{0, 0.05, 0.1\right\}, &
\zeta \in \left\{0, 0.5, 1\right\}, \\
\end{array}
$$
where $\theta=0.2$ (i.e., a variation of $\pm 20\%$). The middle levels are
the values considered when the parameters are fixed.

\begin{figure}[!htb]
\begin{subfigure}[b]{0.5\textwidth}
\centering
\includegraphics[width=\textwidth]{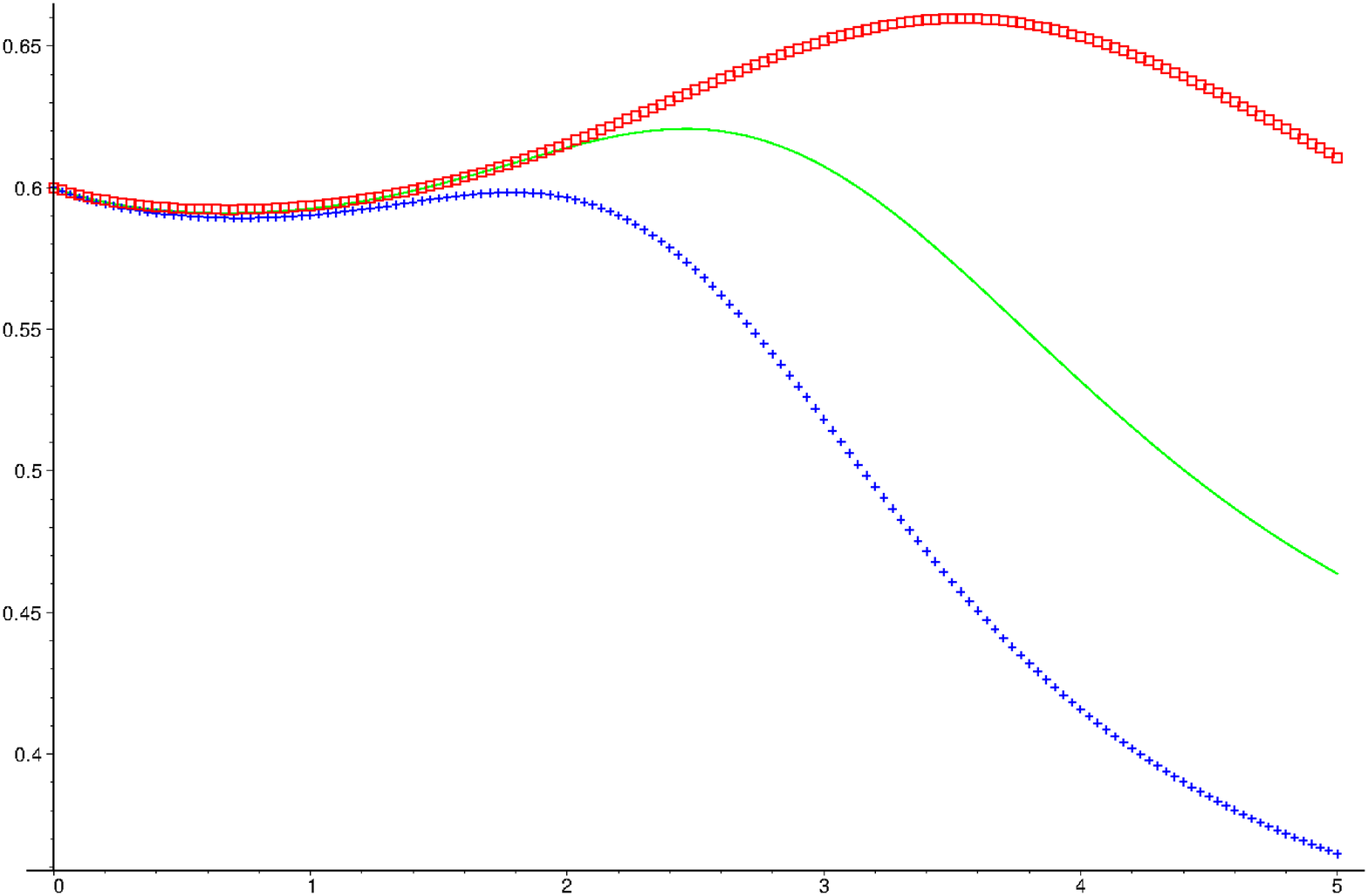}
\SFcaption{$L_A(t) \in [0.35, 0.66]$ vs $t \in [0, 5]$}
\end{subfigure}
\begin{subfigure}[b]{0.5\textwidth}
\centering
\includegraphics[width=\textwidth]{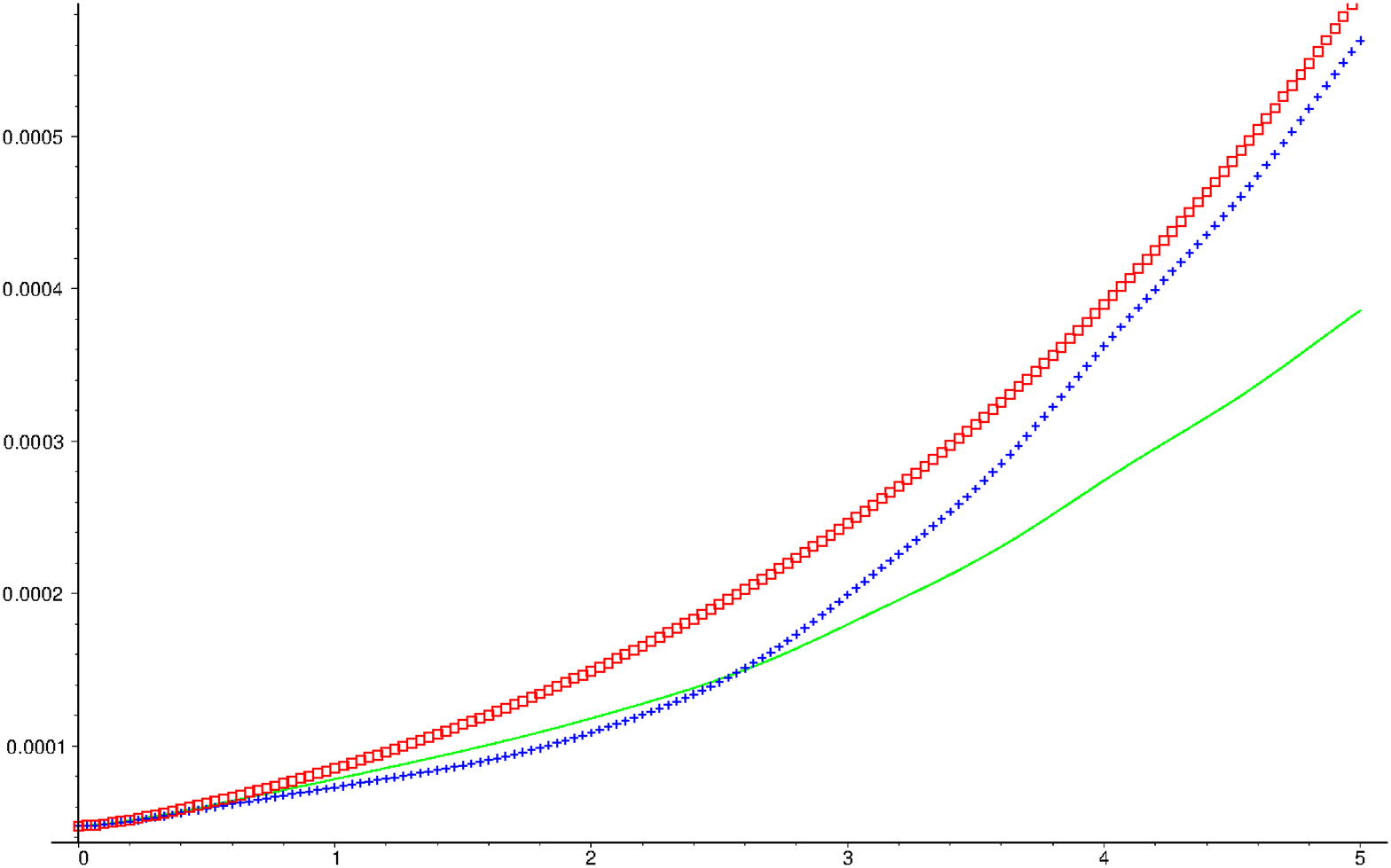}
\SFcaption{$I_G(t) \in [0, 0.0006]$ vs $t \in [0, 5]$}
\end{subfigure}
\caption{$L_A(t)$ when varying $\beta^A$ and $I_G(t)$ when varying
$\zeta$ (box: smaller level, solid: middle level, cross: higher level).}
\label{LAGIZETA}
\end{figure}

Considering that system~\myref{eqPA}--\myref{eqPGamma} has $15$ relevant state-space variables and
we are perturbing $8$ parameters (with $3$ levels), even with overlapping of the
levels on the same graphic, such analysis implies the study of $360$ functions
aggregated in $120$ graphics. We want to quantify and describe the qualitative
behavior and difference between the evolutions, when comparing the different
levels. Additionally, a direct visual interpretation of the plots may be biased
since the plots are not in the same scale, which may give a quite erroneous
filling of disparity between functions when, in fact, the difference may be in
a small amount, e.g., see Fig.~\ref{LAGIZETA}. To deal with such issues,
in a precise and normalized way, we considered the following procedure.

Let $F_{Y,P,1}(t), F_{Y,P,2}(t), F_{Y,P,3}(t)$ be the evolution functions
associated to one of the state-variables
$$
Y \in \{S_A, S_C, S_G, P_A, P_C, P_G, I_A, I_C, I_G, L_A, L_C, L_G, T_A, T_C, T_G\}
$$
and to one of the three variation levels of a parameter
$P\in\{\beta^A, \beta^C, \beta^G, k^C, \phi_T^G, a^A, a^B, \zeta\}$.
Let $\mathcal{T}>0$ denote the total time of simulation. Define
$$
\vartheta(t)= \frac{1}{2}\left(\max_{i\in\mathcal{L}}{F_{Y,P,i}(t)}
+\min_{i\in\mathcal{L}}{F_{Y,P,i}(t)}\right)
\quad\mbox{ andÂ }\quad
\varrho(t)= \frac{1}{4}\left(\max_{i\in\mathcal{L}}{F_{Y,P,i}(t)}
-\min_{i\in\mathcal{L}}{F_{Y,P,i}(t)}\right)^2
$$
for $t\in [0,\mathcal{T}]$ and $\mathcal{L}=\{1,2,3\}$. We divide the analysis
of the graphics, like in Fig.~\ref{LAGIZETA}, in three regions of time:
\textit{beginning} for $t\in \mathcal{B}=[0,\frac{1}{3}\mathcal{T}]$;
\textit{middle} when $t\in\mathcal{M}=[\frac{1}{3}\mathcal{T},\frac{2}{3}\mathcal{T}]$;
and \textit{end} when $t\in\mathcal{E}=[\frac{2}{3}\mathcal{T},\mathcal{T}]$.
The time set for the complete graph is denoted by $\mathcal{A}=[0, \mathcal{T}]$.
Hence, we define
$$
\xi_\mathcal{S}=\frac{\int_\mathcal{S}\varrho(s)\,ds}{\int_0^{\mathcal{T}}
\vartheta(s)\,ds} \:\mbox{ with }\:
\mathcal{S}\in\left\{\mathcal{B},\mathcal{M},\mathcal{E}\right\}.
$$
It is clear, from the linearity of the integral, that
$\xi_\mathcal{A}=\xi_\mathcal{B}+\xi_\mathcal{M}+\xi_\mathcal{E}$.
To understand what $\xi_\mathcal{A}$ measures, consider the hypothetical
situation where $F_{Y,P,1}(t)\equiv m+\theta$, $F_{Y,P,2}(t)\equiv m$, and
consider $F_{Y,P,3}(t)\equiv m-\theta$ for some $m\in\bkR$ and $\theta>0$. Then,
$$
\varphi(t)\equiv m,\:\:\varrho(t)\equiv \theta^2
\quad\Rightarrow\quad \xi_\mathcal{A}=\frac{\theta^2}{m}.
$$
So, although different, $\xi_\mathcal{A}$ is somehow similar to the variance
over the average, which gives an indication of how much the functions are spread
from the average value (between them in each instant of time). The definition of
$\xi_\mathcal{A}$ is also invariant to scale factors, which is quite useful
to eliminate erroneous interpretations of graphics, that may happen without
such measuring tools.

For the qualitative description of the variability of the evolution functions,
we introduced the following tagging notation based on concrete specifications:
\begin{itemize}
\item[1.] (cases $\mathcal{A}_{--}$, $\mathcal{A}_{+-}$, $\mathcal{A}_{++}$)
if $\max(\xi_{\mathcal{B}}, \xi_{\mathcal{M}}, \xi_{\mathcal{E}})<0.4$;
\item[2.] (cases $\mathcal{B}_{--}$, $\mathcal{B}_{+-}$, $\mathcal{B}_{++}$)
if $\mathcal{S}\neq\mathcal{A}$ and $\max(\xi_{\mathcal{B}}, \xi_{\mathcal{M}},
\xi_{\mathcal{E}})=\xi_{\mathcal{B}}$;
\item[3.] (case $\mathcal{M}_{--}$, $\mathcal{M}_{+-}$, $\mathcal{M}_{++}$)
if $\mathcal{S}\neq\mathcal{A}$ and $\max(\xi_{\mathcal{B}}, \xi_{\mathcal{M}},
\xi_{\mathcal{E}})=\xi_{\mathcal{M}}$;
\item[4.] (cases $\mathcal{E}_{--}$, $\mathcal{E}_{+-}$, $\mathcal{E}_{++}$)
if $\mathcal{S}\neq\mathcal{A}$ and $\max(\xi_{\mathcal{B}}, \xi_{\mathcal{M}},
\xi_{\mathcal{E}})=\xi_{\mathcal{E}}$;
\item[5.] (cases $\mathcal{S}_{--}$ with $\mathcal{S}\in\left\{\mathcal{B},
\mathcal{M},\mathcal{E},\mathcal{A}\right\}$) if $\xi_{\mathcal{A}}< 0.01$;
\item[6.] (cases $\mathcal{S}_{+-}$ with $\mathcal{S}\in\left\{\mathcal{B},
\mathcal{M},\mathcal{E},\mathcal{A}\right\}$) if it is not $\mathcal{S}_{--}$
and $\xi_{\mathcal{A}}< 0.25$;
\item[7.] (cases $\mathcal{S}_{+-}$ with $\mathcal{S}\in\left\{\mathcal{B},
\mathcal{M},\mathcal{E},\mathcal{A}\right\}$)
if it is not $\mathcal{S}_{--}$ and $\mathcal{S}_{+-}$.
\end{itemize}
If $\xi_{\mathcal{A}}<0.1$, then we consider that the variation is not
numerically significative, so it is not discussed. Table~\ref{Dvalues}
resumes the sensitivity analysis, where the only tag behaviors that appear are
$\mathcal{B}_{+-}$, $\mathcal{M}_{+-}$, $\mathcal{E}_{+-}$, and $\mathcal{E}_{++}$.
% ------------------------------------------------
\begin{table}[!htb]
\centering
\begin{tabular}{| c | c | c | c | c | l |}
\hline
\ & {\scriptsize $\mathcal{B}_{+-}$} & {\scriptsize $\mathcal{M}_{+-}$}
& {\scriptsize $\mathcal{E}_{+-}$} & {\scriptsize $\mathcal{E}_{++}$}
& {\scriptsize $\xi_\mathcal{A}$ values, respectively}\\
\hline
{\scriptsize $S_A$} & {\scriptsize $a^B$}  & {\scriptsize $\beta^A$}
& {\scriptsize $$} & {\scriptsize $$} & {\scriptsize $\numD{1.17e-01},
\numD{2.64e-01}$}  \\
{\scriptsize $P_A$} & {\scriptsize $$}  & {\scriptsize $$} & {\scriptsize $a^B$}
& {\scriptsize $\beta^A$}  & {\scriptsize $\numD{2.34e-01}, \numD{3.79e-01}$}\\
{\scriptsize $I_A$} & {\scriptsize $$}  & {\scriptsize $$} & {\scriptsize $a^B$}
& {\scriptsize $\beta^A$}  & {\scriptsize $\numD{2.28e-01}, \numD{4.23e-01}$}\\
{\scriptsize $L_A$} & {\scriptsize $$}  & {\scriptsize $$}
& {\scriptsize $\beta^A$} & {\scriptsize $$} & {\scriptsize $\numD{1.26e-01}$} \\
{\scriptsize $T_A$} & {\scriptsize $$}  & {\scriptsize $$}
& {\scriptsize $a^B$} & {\scriptsize $\beta^A$}
& {\scriptsize $\numD{2.18e-01}, \numD{4.58e-01}$}\\
{\scriptsize $S_C$} & {\scriptsize $$}  & {\scriptsize $$}
& {\scriptsize $a^A$} & {\scriptsize $$}  & {\scriptsize $\numD{1.33e-01}$}\\
{\scriptsize $P_C$}  & {\scriptsize $$}  & {\scriptsize $$}
& {\scriptsize $k^C, a^B$} & {\scriptsize $\beta^A, \beta^C, a^A, \zeta$}
& {\scriptsize $\numD{2.47e-01}, \numD{1.82e-01}, \numD{3.78e-01},
\numD{2.84e-01}, \numD{8.43e-01}, \numD{1}$}\\
{\scriptsize $I_C$} & {\scriptsize $$}  & {\scriptsize $$}
& {\scriptsize $\beta^C, k^C, a^B$} & {\scriptsize $\beta^A, a^A, \zeta$}
& {\scriptsize $\numD{1.81e-01}, \numD{3.72e-01}, \numD{1.84e-01},
\numD{3.76e-01}, \numD{8.72e-01}, \numD{9.87e-01}$}\\
{\scriptsize $L_C$} & {\scriptsize $$}  & {\scriptsize $$}
& {\scriptsize $a^A$} & {\scriptsize $$} & {\scriptsize $\numD{2.43e-01}$} \\
{\scriptsize $T_C$} & {\scriptsize $$}  & {\scriptsize $$}
& {\scriptsize $\beta^C, a^B$} & {\scriptsize $\beta^A, a^A, \zeta$}
& {\scriptsize $\numD{1.39e-01}, \numD{1.73e-01}, \numD{3.97e-01},
\numD{9.28e-01}, \numD{9.74e-01}$}\\
{\scriptsize $S_G$} & {\scriptsize $$}  & {\scriptsize $a^A$} & {\scriptsize $$}
& {\scriptsize $\zeta$} & {\scriptsize $\numD{3.95e-01}, \numD{2.88e-01}$} \\
{\scriptsize $P_G$} & {\scriptsize $$}  & {\scriptsize $$}
& {\scriptsize $\beta^A, \phi_T^G, a^A$} & {\scriptsize $\beta^G, k^C, \zeta$}
& {\scriptsize $\numD{1.80e-01}, \numD{1.54e-01}, \numD{2.26e-01},
\numD{7.64e-01}, \numD{4.21e-01}, \numD{3.90e-01}$}\\
{\scriptsize $I_G$} & {\scriptsize $$}  & {\scriptsize $$}
& {\scriptsize $\beta^A, \beta^G, \phi_T^G, \zeta$} & {\scriptsize $k^C, a^A$}
& {\scriptsize $\numD{1.97e-01}, \numD{6.53e-01}, \numD{1.97e-01},
\numD{2.15e-01}, \numD{5.15e-01}, \numD{3.75e-01}$}\\
{\scriptsize $L_G$} & {\scriptsize $$}  & {\scriptsize $a^A$}
& {\scriptsize $\beta^G$} & {\scriptsize $\zeta$}
& {\scriptsize $\numD{2.38e-01}, \numD{9.46e-02}, \numD{2.59e-01}$}\\
{\scriptsize $T_G$} & {\scriptsize $$}  & {\scriptsize $$}
& {\scriptsize $\beta^A, \phi_T^G$} & {\scriptsize $k^C, a^A, \zeta$}
& {\scriptsize $\numD{2.02e-01}, \numD{1.72e-01}, \numD{2.82e-01},
\numD{4.56e-01}, \numD{1.94e-01}$}\\ \hline
\end{tabular}
\caption{Qualitative sensitivity analysis.}
\label{Dvalues}
\end{table}
% ---------------------------------------
Table~\ref{Dvalues} is quite explanatory and shows relations between parameter
perturbations and epidemiological compartments, in a mathematically precise
and rather simple visual representation way. The variation of some parameters
just gives the expected behavior, which shows that the proposed model is suitable
for the situation under study. On the other hand, it also shows that some parameters
that a priori we do not give much attention, as the distribution of persons between
(G) and (C) (i.e., $\zeta$), play an important role in TB spread.

% ------------------------------------------

\section{Conclusions}
\label{sec:conclusions}

In this paper, we propose and analyze a new mathematical model for TB transmission
that considers internal transfer of individuals. As a case-study, we consider a
situation with three populations, namely, Angola (a country with high TB incidence),
people living in a semi-closed community of Angola natives, and other persons living
in Portugal (a country with low TB incidence). Each of the previous subsystems
is divided into five epidemiological categories, which follow the TB transmission
dynamics found in \cite{TBportugalGomesRodrigues}.

For the analysis and verification of the results presented in this paper,
we developed a software tool, so-called \textit{sDL} \cite{SDL}, that combines in the
same framework the power of pre-processing systems (as \textit{m4} \cite{m4}
and \textit{cpp} \cite{cpp}), a logical verification tool for classical
and hybrid systems (as \textit{SMT} \cite{smt} or \textit{KeYmaera} \cite{Platzer}),
a computer algebra system (as \textit{Maple} \cite{maple}), and a numerical
computing language (as \textit{Matlab} \cite{matlab}). The pre-processing systems
allow the existence of a unique and general file, where constants and ODEs
are defined in two hierarchical levels, in order to be used across all tools.
The verification tool and the computer algebra system allowed to test the
validity of some assumptions and verify the correctness of analytic/algebraic
formulae. As expected, the numerical computing language allowed to do the numeric
simulations and generate the corresponding graphics. Considering the potential
of the software tool \textit{sDL}, in a forthcoming publication, we intend to
study real situations that are modeled by pure hybrid model systems, e.g.,
transmission coefficients that are discontinuous functions varying with
climate and season conditions.

Simulations and sensitivity analysis show that variations of the transmission
coefficient on the origin country has a big influence on the number of infected
(and infectious) individuals on the community and the host country. This enforce
the importance of an additional effort to treat TB and improve health conditions
in countries with high TB incidence, since they remarkably affect (in long term)
the health of individuals on other countries. As expected, an increment on the
flux of individuals moving from areas of lower TB incidence to areas of higher
TB incidence reduces the global reproduction number and an increment in the flux
of individuals moving from areas of high TB incidence to areas of lower TB
incidence increases the global reproduction number, but also introduce
modifications in the evolution of each disease category that is not linearly
proportional to flux rate. From the community point of view, it is better to
have some moderate exchange of persons with the high incidence TB region.
Seasonality distribution of persons traveling between Angola and Portugal
has an important impact in the number of infected (and infectious) individuals
in the community.

The main conclusion is that, contrary to some beliefs, the existence of a
community of immigrants coming from a high incidence TB area seems to be
convenient in a global point of view, as well as for the host country,
in order to better control TB spread. On the other hand, it does not affect
the TB incidence in the origin country of the immigrant community.
By nonexistence of the community of immigrants we mean the situation where
the individuals traveling are spread uniformly on the host country. As shown
above, a key parameter in such analysis is the percentage of persons traveling
from the high incidence TB area that will stay in the community. Such parameter
has an optimal value for TB control, in the sense of minimizing
the global reproduction number, that is near to $47\%$.
The obtained results are valid under the hypothesis of
a semi-closed community. Further studies are necessary
for the situation without any flux restrictions.

% ------------------------------------------

\bigskip

\noindent \textbf{Acknowledgments}.
{\small Work partially supported by Portuguese funds through the Center for Research
and Development in Mathematics and Applications (CIDMA) and the Portuguese
Foundation for Science and Technology (FCT), within project UID/MAT/04106/2013.
Rocha is also supported by the FCT project ``DALI -- Dynamic logics for cyber-physical 
systems: towards contract based design'' 
with reference  P2020-PTDC/EEI-CTP/4836/2014;
Silva by the FCT post-doc fellowship SFRH/BPD/72061/2010;
Silva and Torres by project TOCCATA, reference PTDC/EEI-AUT/2933/2014, funded by Project 
3599 -- Promover a Produ\c{c}\~ao Cient\'{\i}fica e Desenvolvimento
Tecnol\'ogico e a Constitui\c{c}\~ao de Redes Tem\'aticas (3599-PPCDT)
and FEDER funds through COMPETE 2020, Programa Operacional
Competitividade e Internacionaliza\c{c}\~ao (POCI), and by national
funds through FCT. The authors are grateful to two referees 
for useful comments and suggestions.}

% ------------------------------------------

\small

% ------------------------------------------

% ------------------------------------------

\end{document}